\documentclass[]{aastex}
  
   \usepackage[]{graphicx}
  \usepackage{caption}
\usepackage[]{natbib}
\usepackage{url}

\usepackage{multirow}
\usepackage{epstopdf}
\shorttitle{A devoted mission to characterize exoplanets}
 \shortauthors{Tessenyi M., Ollivier M., Tinetti G., Beaulieu J.P. \emph{et al.}}
\begin{document}
\title{Characterising the Atmospheres of Transiting Planets with a Dedicated Space Telescope}
\author{M. Tessenyi\altaffilmark{1,2},
M. Ollivier\altaffilmark{3,4},
G. Tinetti\altaffilmark{1},
J.P. Beaulieu\altaffilmark{2},
V. Coud\'e du Foresto\altaffilmark{5},
}
\author{T. Encrenaz\altaffilmark{5},
G. Micela\altaffilmark{6},
B. Swinyard\altaffilmark{7,1},
I. Ribas\altaffilmark{8},
A. Aylward\altaffilmark{1},
J. Tennyson\altaffilmark{1},
}
\author{M.R. Swain\altaffilmark{9}, A. Sozzetti\altaffilmark{10}, G. Vasisht\altaffilmark{9} \& P. Deroo\altaffilmark{9} }

 \altaffiltext{1}{University College London, Department of Physics and Astronomy, Gower Street, London WC1E 6BT, UK}
\altaffiltext{2}{Institut d'Astrophysique de Paris, CNRS, UMR7095, Universit\'e Paris VI, 98bis Boulevard Arago, PARIS, France}
\altaffiltext{3}{Universit\'e Paris-Sud, IAS UMR8617, ORSAY F-91405}
\altaffiltext{4}{CNRS, ORSAY F91405}
\altaffiltext{5}{Observatoire de Paris, LESIA, Meudon, France }
\altaffiltext{6}{INAF - Osservatorio Astronomico di Palermo, Piazza del Parlamento 1, 90134 Palermo, Italy}
\altaffiltext{7}{RAL Space, STFC-Rutherford Appleton Laboratory, Harwell Campus, Chilton, Didcot, Oxon, OX11 0QX, UK}
\altaffiltext{8}{Institut de Ciencies de l'Espai (CSIC-IEEC), Campus UAB, 08193 Bellaterra, Spain}
\altaffiltext{9}{Jet Propulsion Laboratory, California Institute of Technology, 4800 Oak Grove Drive, Pasadena, California 91109-8099, USA}
\altaffiltext{10}{INAF - Osservatorio Astronomico di Torino, Strada Osservatorio 20, 10025 Pino Torinese (TO), Italy }
\begin{abstract}
Exoplanetary science is one of the fastest evolving fields of today's astronomical research, continuously yielding unexpected and surprising results. Ground-based planet-hunting surveys together with dedicated space missions such as Kepler and CoRoT, are delivering an ever-increasing number of exoplanets, now numbering at over 690, and ESA's GAIA mission will escalate the exoplanetary census into the several thousands.
The next logical step is the characterisation of these new worlds: what is their nature? Why are they as they are? 
The use of the Hubble and Spitzer Space Telescopes to probe the atmospheres of transiting hot, gaseous exoplanets has opened perspectives unimaginable even just ten years ago, demonstrating that it is indeed possible with current technology to address the ambitious goal of characterising the atmospheres of these alien worlds.
These successful measurements have however also shown the difficulty of understanding the physics and chemistry of these exotic environments when having to rely on a limited number of observations performed on a handful of objects. 

To progress substantially in this field, a dedicated facility for exoplanet characterisation, able to observe through time and over a broad spectral range a statistically significant number of planets, will be essential.
Additionally, the instrument design (e.g. detector performances, photometric stability, etc.) will be tailored to optimise the extraction of the astrophysical signal.
In this paper, we analyse the performance and trade-offs of a 1.2/1.4\,m space telescope for exoplanet transit spectroscopy from the visible to the mid IR.

We present the signal-to-noise ratio as a function of integration time and stellar magnitude/spectral type for the acquisition of spectra of planetary atmospheres for a variety of scenarios: hot, warm, and temperate planets, orbiting stars ranging in spectral type from hot F to cooler M dwarfs.
Our results include key examples of known planets (e.g. HD 189733b, GJ 436b, GJ 1214b, and Cancri 55 e) and simulations of plausible terrestrial and gaseous planets, with a variety of thermodynamical conditions. We conclude  that even most challenging targets, such as super-Earths in the habitable-zone of late-type stars, are within  reach of a M-class, space-based spectroscopy mission.

\end{abstract}
\maketitle

\label{firstpage}
\section{Introduction}
The science of extra-solar planets is one of the most rapidly changing areas of astrophysics and since 1995 the number of planets known has increased by almost two orders of magnitude. ÊA combination of ground-based surveys and dedicated space missions has resulted in 690-plus planets being detected \citep{schneider_extrasolar_2010}, and over 1200 that await confirmation \citep{kepler}. ÊNASA's Kepler mission has opened up the possibility of discovering Earth-like planets in the habitable zone around some of the 100,000 stars it is surveying during its 3 to 4-year lifetime. The new ESA's Gaia mission is expected to discover thousands of new planets around stars within 200 parsecs of the Sun \citep{gaia,gaia1}.
Meanwhile, transit and combined light methods have allowed the characterisation of the atmosphere of a few hot large bodies close to their star using current space telescopes, 
\citep[e.g.][]{charbonneau_detection_2001,harrington_phase-dependent_2006,crossfield2010,knutson_using_2007,tinetti_2007,tinetti_2010,beaulieu_primary_2007,beaulieu_2009,swain2008a,swain2008b,swain2009a,swain2009b,grillmair_strong_2008,stevenson2010} and ground based telescopes \citep{redfield_sodium_2007,snellen_ground-based_2008,swain_ground-based_2010,snellen2010,waldmann2011}. Transiting hot super-Earths, while being very interesting targets since they are absent from our Solar System, are within reach  with current telescopes, e.g. GJ 1214b \citep{charbonneau_super-earth_2009,bean2010}, and Cancri 55 e \citep{winn2011}. \\
The next generation of ground-based telescopes and the James Webb Space Telescope will have a noticeably larger collecting area compared to current facilities,
allowing them to probe fainter targets in the future.
 However, these  facilities are built for the larger astrophysics community and are not necessarily optimised for exoplanet characterisation. The investigation of exoplanetary atmospheres requires
a dedicated space mission that is fine-tuned to this purpose. Such a mission should be able not only to simultaneously capture the spectral signatures over a broad wavelength region to reveal the chemical and dynamical processes of the atmosphere, but also have enough time to observe many systems repeatedly. These systems should  
include the dimmer planets that approach the size of Earth, hence the instrument design should  be optimised to eliminate systematic errors.
\\ 
In this paper, we consider the possibilities offered by --- and the trade-offs of --- a 1.2 to 1.4 m  space based  telescope capable of performing spectroscopy from
the visible down to  the Mid-IR.  
A similar mission concept has most recently been selected for an assessment study by ESA, under the name Exoplanet Characterisation Observatory (EChO)\footnote[1]{http://sci.esa.int/echo/} \citep{tinetti2011}.
\section{Methods}
\subsection{Classification of planetary atmospheres}
\label{sec:spectra}
We classified the planetary atmospheres according to equilibrium temperatures and sizes, in particular:
three classes of atmospheric temperatures: `\emph{Hot}' (800-2000 K), `\emph{Warm}' (350-800 K) and `\emph{Habitable}' (250-350 K) and three types of planetary sizes: \emph{Jupiter-like, Neptune-like} and \emph{super-Earth} (see Table \ref{tab:tempsize}). Planets with `\emph{Cold}' temperatures (200 K or below) are not studied in this paper.

\begin{table}[h!]
\small
\centering
\begin{tabular}{l c c c}
\hline
Temperature/Size &  Jupiter-like &  Neptune-like &  super-Earth \\
\hline
Hot &   \textbf{HJ} &   HN &  \textbf{HSE}  \\
Warm & WJ & \textbf{WN}  & \textbf{WSE} \\
HZ & \textbf{HZ-J} & HZ-N  & \textbf{HZ-SE}  \\
\hline
\end{tabular}
\caption{ \footnotesize Subdivision of planetary atmospheres according to temperature and planetary-size. The difficulty in the observations increases from left to right and from top to bottom. All categories in bold are studied in detail in section \ref{results}, results for the three other categories can be extrapolated.  }
\label{tab:tempsize}
\end{table}
Super-Earths are expected to be between 1 and 10 $M_{\oplus}$, in this paper we assume a 5 Earth-mass body for our calculations, with a radius of 1.6-1.8 $R_\oplus$  \citep{grasset_study_2009}.
By comparing the Earth's cross section:   $\sigma_\oplus = \pi \cdot R_\oplus^2$ to those of the super-Earth, Neptune-like and Jupiter-like planets, we obtain:
\begin{equation}
\sigma_{SE} \sim 3  \, \sigma_\oplus; \qquad \sigma_{N} \sim 25 \,  \sigma_\oplus; \qquad \sigma_{J} \sim 100  \sigma_\oplus
\end{equation}
For transit and combined-light observations (transiting and non-transiting planets), the important parameter is  the ratio between the planetary and the stellar cross sections, $\kappa$, obtainable by measuring the transit depth:
\begin{equation}
\kappa = \sigma_p/\sigma_* 
\label{eq:kappa}
\end{equation}
\begin{table}[h!]
\footnotesize
\centering
\begin{tabular}{l l c l c c c c}
\hline
Star type &    Temp. (K) & Radius ($ R_\odot$)  &  $\sigma_{*}$ ($\sigma_\odot$) & & $\kappa_{Jup.}$ ($\kappa_J$)  &  $\kappa_{Nept.}$ ($\kappa_J$) & $\kappa_{SE}$ ($\kappa_J$) \\
\hline
F3V  &   6740K &  1.56  &   $\sigma_{F3} \sim 2.4 \, $ & & $\sim$ 0.5 & $\sim$ 0.05 & $\sim$0.01  \\
G2V  & 5800K   &  1    &  $\sigma_{G} = \, \sigma_\odot$ & & 1 & $\sim$ 0.1 & $\sim$ 0.02 \\
K1V &  4980K  & 0.8  & $\sigma_{K1} \sim 0.6 \, $ &  & $\sim$ 2 & $\sim$ 0.2 & $\sim$ 0.03\\
M1.5V & 3582K  & 0.42 & $\sigma_{M1.5} \sim 0.18$ & & $\sim$ 6 & $\sim$ 0.7 & $\sim$ 0.1 \\
M3.5V & 3376K & 0.26 & $\sigma_{M3.5} \sim 0.07 \,$ & & $\sim$ 15 & $\sim$ 2 & $\sim$ 0.3  \\
M4.5V & 3151K  & 0.17  & $\sigma_{M4.5} \sim 0.03 \,$ & & $\sim$ 35 & $\sim$ 4 & $\sim$ 0.7 \\
M6V & 2812K  & 0.12 & $\sigma_{M6} \sim 0.01 \, $ & & $\sim$ 70 & $\sim$ 9 & $\sim$ 2 \\
\hline
\end{tabular}
\caption{\footnotesize Cross  section $\sigma_{*}=\pi R_{*}^2$ for different stellar types and corresponding $\kappa$ values for the three planet sizes considered: Jupiter-like, Neptune-like and super-Earth. The reader can note that super-Earths in the orbit of late M stars have a similar ratio $\kappa$ to a Jupiter in the orbit of a Sun-like star.}
\label{tab:star}
\end{table}
This parameter changes significantly for different  planet/star  types. In  Table \ref{tab:star}, we give  $\sigma_{*}$ for few key stellar types, along with the cross-section ratio value $\kappa$ for the three planetary types considered in this paper, expressed as a factor of $\kappa_{Jupiter}$ ($\kappa_{Jup.}\sim 100 \sigma_\oplus / \sigma_\odot$). A Jupiter-sized planet orbiting a Sun-like star and a super-Earth orbiting a M4.5 dwarf will both have a similar cross-section ratio $\kappa \sim \kappa_{Jup.}$, observable with small, ground based telescopes.
\subsection{Primary Transit method}
\label{sec:primary}
A primary transit occurs when a planet passes in front of its parent star with respect to our line of sight. By subtracting the ``in-transit" stellar flux  from the ``out-of transit",
we can measure directly the parameter $\kappa$ as described in eq. \ref{eq:kappa}, hence the planetary radius in units of stellar radii. If we repeat the observation of 
$\kappa$ at different wavelengths, we can infer the presence or absence of an atmosphere as well as retrieve the main atmospheric components \citep{seager_sasselov,brown}.
The spectral absorption of the planetary atmosphere, while in transit, is measured from the transmission spectrum obtained. 
For key examples of planetary cases, we use synthetic models  fitting the existing observations or we extrapolate from our knowledge of the planets in the Solar System. The models were calculated with  the line-by-line radiative transfer code described in \citep{tinetti_2007a,tinettiRS2011}, with updated line-lists from \citet{barber,byte,rothman,tashkun}. 
For feasibility studies, we adopt as well a more heuristic estimate of the atmospheric contribution rather than these detailed simulations.
In particular, the amount of light passing through the atmosphere of the planet will cross a small annulus:
\begin{equation}
\label{eq:primarysignal}
\frac{2 R_p \pi \Delta z}{\pi {R_\star}^2} = \frac{2 R_p \Delta z}{{R_\star}^2}  
\end{equation}
where $R_p$ is the radius of the planet, $R_\star$ the radius of the star and $\Delta z$ the height of the atmosphere.
From observations $\Delta z = nH$, with typically $n\sim5$, depending on the spectral resolution and wavelength. $H$ is the scale height defined by:
\begin{equation}
\label{eq:scaleheight}
H = \frac{k\,T}{\mu\,g}
\end{equation}
where $k$ is the Boltzmann constant, $g$ is the gravity acceleration and  $\mu$ the mean molecular mass of the atmosphere.\\
From the scale height expression (\ref{eq:scaleheight}), it is clear that the hotter and lighter the atmosphere is, the easier it is to observe it with this technique. Also, dense objects such as telluric bodies will have a higher value for $g$ and consequently a more compact atmosphere.
For example, hot Jupiters have high temperatures, low mean molecular mass of the atmosphere ($\mu\sim2$~amu for hydrogen-rich atmospheres) and relatively low density. Their scale height can easily reach 500 km.
By contrast, the Earth's temperature is colder ($\sim$280 K),
$\mu$  is \nolinebreak $\sim28$~amu and the bulk composition is denser. As a result the scale height is compacted to \nolinebreak $\sim8$ km.\\
As explained in \S 2.1, we are interested here in 3 main classes of planets: gas-giants, Neptune-like and super-Earths. While gas giants and Neptunes are optimal targets for primary transit observations,
in general super-Earths are the least favourable, unless they transit an M star and host a relatively hot and light atmosphere (see Table \ref{tab:primaryvssecondary}).
For this reason, in the most general case, super-Earths should be observed with the secondary eclipse technique, as described in the next section.

\subsection{Secondary Eclipse method}
\label{sec:contrast}
\begin{figure}[t]
   \begin{center}
         \includegraphics[width=3.1in]{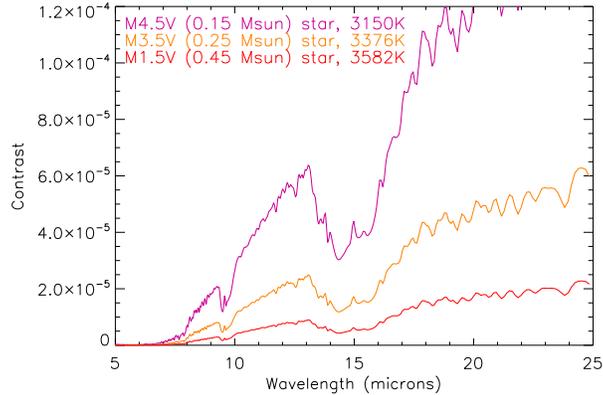}
   \caption{ \small Planet/star flux contrast (eq. \ref{eq:fluxcontrast})  for a super-Earth orbiting different M-type stars (M1.5V, M3.5V and M4.5V). In this example the super-Earth is assumed to have an Earth-like atmosphere. (see section \ref{sec:hzplanets}) }
   \label{fig:ve200all}
   \end{center}
   \end{figure}

\begin{figure}[t]
   \begin{center}
   \includegraphics[width=3.1in]{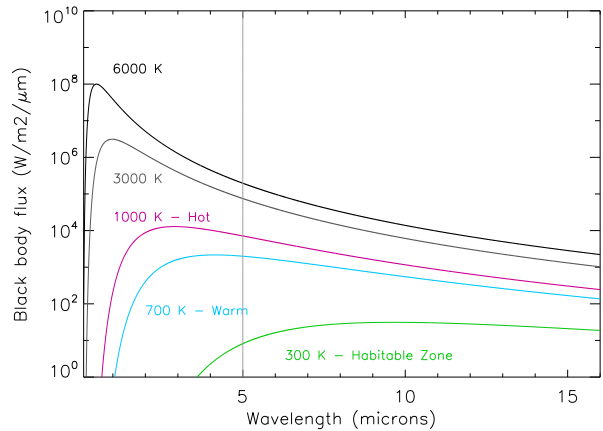}
   \caption{ \small Blackbody curves for effective temperatures of 6000, 3000, 1000, 700 and 300 K.  \newline The radiation emitted by the 300 K body is negligible at $\lambda$ shorter than 5 $\mu$m.}
   \label{fig:planck}
   \end{center}
   \end{figure}

A complementary technique to the primary transit, is the so called secondary eclipse. This method relies on the possibility of observing the star alone when the planet is passing behind it, so we can effectively subtract the stellar contribution from the star+planet system. 
 In practice, we measure the flux emitted and/or reflected by the planet in units of the stellar flux:
\begin{equation}
\label{eq:fluxcontrast}
F_{II} (\lambda) = {\left( \frac{R_p}{R_\star} \right)}^2 \frac{F_p(\lambda) }{F_\star(\lambda)} = \kappa \, \frac{F_p(\lambda) }{F_\star(\lambda)} 
\end{equation}
where $F_p$  and $F_\star$ are the planetary and stellar spectra, and $\kappa$ was defined in \S 2.1. It is clear from this equation that both the relative size  planet/star  (the parameter $\kappa$) and the relative
temperature are key parameters for secondary eclipse measurements. 
Here we use  synthetic models to represent key examples of exoplanets \citep{tinetti_faraday}. The emitted/reflected spectra were generated with the line-by-line radiative transfer codes described in
\citet{tinetti_mars,tinetti_infrared_2006}, with updated line-lists from \citet{barber,byte,rothman}. For these key cases, the stellar spectra are either observed or modelled    \citep{kurucz_solar_1995}.
In Fig. \ref{fig:ve200all} we show how  $F_{II} (\lambda)$  changes for a given planet as a function of star type. In this example, we chose a super-Earth with an Earth-like atmosphere orbiting a selection of known M-dwarfs: clearly late M stars offer the best planet/star contrast.
\paragraph{Infrared observations} For feasibility studies in the IR, we approximate the planetary and stellar spectra in eq. \ref{eq:fluxcontrast} with two Planck curves at temperature $T_p$ and $T_\star$, with $T_p$ being the day-side temperature of the planet. While this approximation is not accurate enough to model specific examples, it is helpful to estimate the general case.
\begin{equation}
F_{II} (\lambda) \sim  \kappa \, \frac{B_p(\lambda, T_p) }{B_\star(\lambda, T_\star)} 
\label{eq:fluxbb}
\end{equation}
In Fig.  \ref{fig:planck} we show the Planck curves for a few bodies at different temperatures. The planet to star flux contrast will clearly be higher for hot planets.  Note that in the IR temperate planets at $\sim$300 K can be observed only at wavelengths longer
than 5 $\mu$m, as they emit a negligible amount of flux at  $\lambda \le 5 \mu$m  (Fig.  \ref{fig:planck}).
\paragraph{Optical observations} 
For observations in the optical, we need to estimate the reflected light from the planet. Eq. (\ref{eq:fluxcontrast}) becomes:
\begin{equation}
\label{vis}
F_{II} (\lambda) = {\left( \frac{R_p}{R_\star} \right)}^2 \frac{F_p(\lambda) }{F_\star(\lambda)} = \kappa \, A \, \zeta \, \frac{R_*^2}{a^2}  \, \frac{F_\star(\lambda) }{F_\star(\lambda)} = \kappa \, A \, \zeta \, \frac{R_*^2}{a^2}
\label{eq:visible}
\end{equation}
where $A$ is the planetary albedo, $\zeta$ is the observed fraction of the planet illuminated and $a$ the semi-major axis.  The closer the planet to its stellar companion and the higher its albedo, the larger the contrast in the optical will be. For planets colder than $\sim 1200$K, the reflected light component is predominant in the optical wavelength range ($\lambda < 0.8 \mu m$). For hotter planets, both equations \ref{eq:fluxbb} and \ref{vis} will bring a contribution (emission and reflection).

\paragraph{Planet Phase Variations and Eclipse Mapping} 

Phase-variations are important in understanding a planet's atmospheric dynamics and the redistribution of absorbed stellar energy from their irradiated day-side to the night-side. 
These observations can only be conducted from space since the typical time scale of these phase variations largely exceeds that of one observing night. 
Phase variations are very insightful both at reflected and thermal wavebands.
 In the infrared case, these kinds of observations are critical to constrain General Circulation Models of exoplanets, of hot gaseous planets in particular. For instance, the infrared $8\mu m$ Spitzer observations of the exoplanet HD189733b have shown the night-side of this hot Jupiter to be only $\sim300K$ cooler than its day-side \citep{Knutson2007}, suggesting an efficient redistribution of the absorbed stellar energy. In addition, towards the optical wavelength regime, an increasing contribution from reflected light is expected \citep{Snellen2009,Borucki2009}.

A great advantage of a dedicated exoplanet mission would be the potential for long campaigns: staring at a known planetary system for a sizable fraction of \citep{Knutson2007, Knutson2009a, Knutson2009b} or an entire orbit \citep{Snellen2009, Borucki2009}, or ---provided the flux calibration is accurate enough--- using multi-epoch observations to obtain a more sparsely sampled phase curve \citep{Cowan2007, crossfield2010}.  At thermal wavelengths this may only be interesting for short-period planets, where the diurnal temperature contrast is high.  Additionally, non-transiting planets open up interesting possibilities to study seasons (eg, \citet{Gaidos2004}). Furthermore, the simultaneous multi-band coverage would make it possible to simultaneously probe the longitudinal temperature distribution as a function of pressure, which would be a very helpful constraint for GCMs.

The potential for using phase variations to study non-transiting systems should also be noted \citep{Selsis2011}.  Non-transiting systems are going to be closer on average than their transiting counterparts.  The challenge is stellar and telescope stability over the orbital time of a planet. 
For planets on circular orbits, thermal phases have limited value because of the inherent degeneracies of inverting phase variations \citep{Cowan2008}, but for eccentric planets, phase variations will be much richer \citep{Langton2008,Lewis2010, Iro2010, Cowan2011}.  As one considers increasingly long-period planets (warm rather than hot) even more of them will be on eccentric orbits because of the weaker tidal influence of the host star.

For the brightest targets, eclipses can also be used as powerful tools to spatially resolve the emission properties of planets. During ingress and egress, the partial occultation effectively maps the photospheric emission region
of the object being eclipsed \citep{Williams2006,Rauscher2007,Agol2010}. Key constraints can be placed  on 3D atmospheric models through repeated infrared measurements.
In this paper, we will focus on the feasibility of primary transits and secondary eclipses. A more detailed and thorough study of the observability of phase variations and eclipse mapping will be the topic of future publications.

\subsection{Comparison between primary and secondary transit techniques}
\label{sec:primaryvssecondary}
The primary and secondary transit techniques are complementary. Transmission spectra in the infrared, from primary transits, are sensitive to atomic and molecular abundances, but less to temperature gradients. 
In comparison, emission spectroscopy allows for detection of molecular species alongside constraining the bulk temperature and vertical thermal gradient of the planet.
Additionally, during the primary transit we can sound the terminator, whereas during the secondary eclipse we can observe the planetary day-side.
\begin{table}[ht]
\centering
\small 
\begin{tabular}{ l  c c c }
  \hline
 & Jupiter & Neptune & super-Earth\\
\emph{star:}  & K & M2.5V & M4V\\
  \hline
Hot  & 0.18  & 0.98  & 0.3 / 0.09 \\
Warm & 0.42  & 2.17  & 0.7 / 0.2  \\
HZ   & 0.9   & 10.4  & 1.2 / 0.3  \\
  \hline
\end{tabular}
\caption{ \footnotesize Primary / secondary eclipse flux ratio for key examples of the planetary classes listed in Table \ref{tab:tempsize}. Numbers $>1$ indicate that the primary transit is more favourable over the secondary, while numbers $<1$ indicate the opposite.
The results are obtained by dividing the atmospheric signals calculated from equations \ref{eq:primarysignal} and \ref{eq:fluxcontrast}, taken at $\sim10 \mu m$ for all presented cases.
For the super-Earth we report two values: a case of an ``ocean planet'' (1.8 $R_\oplus$, \citep{grasset_study_2009}) with water vapour being the main component of the planetary atmosphere, and a telluric planet with CO$_2$ as main atmospheric component (1.6 $R_\oplus$). In the habitable-zone, the ratio for the latter case is less favourable, with 0.3 excluding the possibility of primary transit studies. By contrast, for an ``ocean planet'', the ratio of 1.2 is similar to the ratio for the habitable-zone Jupiter-like planet.
}
\label{tab:primaryvssecondary}
\end{table}

In Table \ref{tab:primaryvssecondary} we present ratios of signal values from primary transit and secondary eclipse observations for the key examples of planetary classes (see Table \ref{tab:tempsize}).
Given that long integration times require the co-adding of multiple transit observations, for the primary case, any systematic difference in the stellar flux could hamper results. For example, spot redistributions over the stellar surface could potentially alter the depth of the transit, and could be a reason of concern for late-type stars since, on average, they can be quite active. In the case of M-type star super-Earths, though, we rely mostly on secondary eclipse observations which are quite immune from effects related to stellar activity, as the planetary signal follows directly from the depth of the occultation without the need to model the stellar surface.

\subsection{Planets orbiting M-type Stars}
\label{sec:stars}
In this section we focus our attention  on M-class stars and their habitable zone. The main reasons to consider them are:
\begin{enumerate}
\item Among the stars  in the solar neighbourhood 90\% are of M-type (e.g. \citet{perryman97}). 
\item The relative small size of M stars (typically between 0.08 - 0.5 $R_\odot$) allows us to probe  planetary sizes down to a few earth masses (see Eq. \ref{eq:fluxcontrast}).
\item
The low effective temperature of the star (2900$<T_{eff}<$3900 K), places the habitable zone (HZ) region closer-in to the star than would be the case for a hotter star. A HZ planet will hence have a short orbital period (see Fig. \ref{fig:star_period}) and a larger number of transit events will be observable within a given time interval than would be the case for a planet in the HZ of hotter (K, G, F) stars.
\item M stars are brightest in the IR (i.e.: more photons impinging onto the detector), where the temperate exoplanets are easier to observe (see Fig. \ref{fig:planck}).
\end{enumerate}
The combination of these  effects brings the prospect of characterising  terrestrial planets in the HZ of main sequence stars within current technology capabilities.
By contrast, it is currently impractical to use the transit technique to observe the atmosphere of terrestrial planets in the HZ of more massive stars, as the orbital period in this case would be very long (e.g. more than 100 days for a K-type star and 300 for a G-type star).
In addition, notice that M dwarf spectra differ significantly from blackbody radiation curves in the visible and near infrared parts of the spectrum, but in the mid-infrared, considered for our habitable-zone targets, the molecular absorptions are less important.
\subsubsection{M-star population }   
    \begin{figure}[ht]
   \begin{center}
   \includegraphics[height=3.3in]{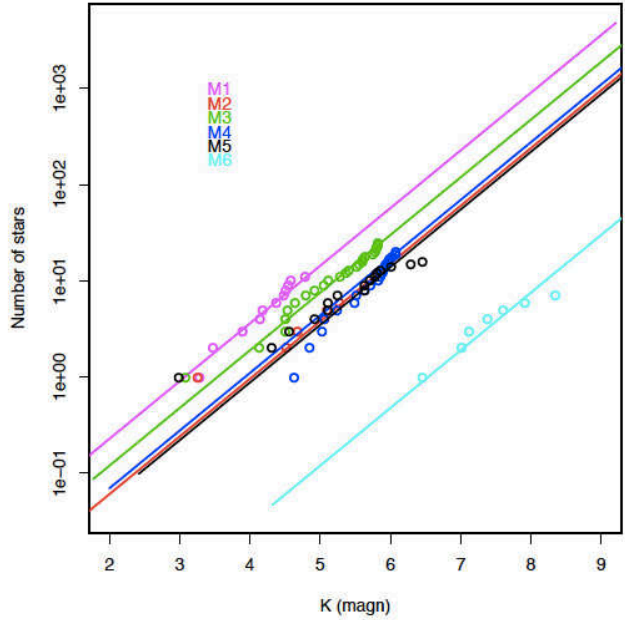}
   \caption{ \small  Expected number of stars out to 10 pc,  for M0-4V and M5-M9V. Dots are stars in K magnitude from the RECONS catalogue and lines represent the expectations, assuming uniform spatial distribution and completeness at 6.6 pc.
These plots suggest that the RECONS catalogue is complete only up to 6.6 pc for the earliest spectral types and up to 4.5-6 pc for the M5-6V sample. There are too few objects in the M7-9V range to say anything about completeness/space density of such objects. }
   \label{fig:recons}
   \end{center}
   \end{figure}
At the time of writing over 25\% of stars in the Sun's near neighbourhood are believed to be missing from star surveys \citep[such as the catalogue by][]{lepine2011}, in part because bright M stars in the infrared are quite faint in the visible, due to a combination of temperature and the presence of molecular and atomic species absorbing in this spectral region. For instance, a M3V star with V=12.30 mag. corresponds to K=7.53 mag., and a M5V star with V=15.01 mag. corresponds to a K=8.40 mag. \citep{delfosse2000}.
For this reason in this paper we use K magnitude rather than V to classify the luminosity of M stars.
The most complete catalogue of late-type nearby stars available
today is the \citet{lepine2011} catalogue, which includes nearly 9000 M dwarfs with magnitude J $< 10$. According to the authors, the catalogue represents $\sim75$ \% of the of the estimated $\sim11,900$ M dwarfs with J$<10$ expected to populate the entire sky.

An evaluation of the number of M stars in a magnitude-limited sample can be derived also
from the analysis of the 100 RECONS nearest star systems \citep{recons100}. Their distribution in distance shows
clearly that, while the M1-4V star sample is evenly distributed within 6.7 pc, the M5-8V sample
is significantly incomplete beyond 4-5 pc (see Fig. \ref{fig:recons}). This analysis supports the hypothesis that a significant number of stars are still missing in catalogues also in the very close solar neighborhood; there needs to be a major effort in the next years in this direction.
\\
Independent estimates of the M star population in the solar neighbourhood were provided by Micela G. (private communication) through colour-colour
diagrams  applied to the 2MASS (Two Micron All Sky Survey) catalogue. This selection might have some contamination from two different sources:
a) Distant giant stars may overlap with the nearby very early M type, where the main sequence and giant
models are still close together.
b) Stars of early spectral type could contaminate the dwarf M-star regime only if highly
reddened by intervening dust. 
\\
NASA's Wide-field Infrared Survey Explorer (WISE) might help with removing the contamination, by providing a survey in four additional IR channels \citep{wise}.
\\
The Gaia mission, in its all-sky astrometric survey, will deliver direct parallax
estimates and spectrophotometry for nearby main-sequence stars down to
R$\sim$20. At the magnitude limit of the survey, distances to relatively bright
M stars out to 20-30 pc will be known with 0.1\%-1\% precision
(depending on spectral sub-type). This will constitute an improvement of
up to over a factor 100 with respect to the typical 25\%-30\% uncertainties
in the distance reported for low-mass stars identified as nearby based on
proper-motion and colour selections \citep[e.g.][]{lepine2011}. Starting with
early data releases around mid-mission, the Gaia extremely precise
distance estimates, and thus absolute luminosities, to nearby late-type
stars will allow us to improve significantly standard stellar evolution models at the
bottom of the main sequence. For transiting planet systems, updated values
of masses and radii of the host stars will be of critical importance.
Model predictions for the radii of M dwarfs show today typical
discrepancies of $\sim15\%$ with respect to observations, and, as shown by
the GJ 1214b example (Charbonneau et al. 2009), limits in the knowledge of
the stellar properties significantly hamper the understanding of the
relevant physical characteristics (density, thus internal structure and
composition) of the detected planets.
Meanwhile, to account for the possibility of errors in current measurements, we provide a variety of stellar temperatures and calculated corresponding star radii with our results. The radii were calculated using isochrones of old ($\sim$ 5bn yr) low-mass stars \citep{baraffe98}, and observational constrains \citep{delfosse2000}. For comparison, based on the simple radius-temperature-luminosity relation considerations, we can infer that estimates of stellar radii, when Gaia parallaxes known to $<1\%$ will become available for nearby red stars, will carry much reduced uncertainties, on the order of 1\%-3\%.
Indeed, the precision in the M dwarf effective temperature estimates from spectroscopy or photometric calibrations (currently, 3\%-5\% at best) will then become the limiting factor in the knowledge of this fundamental quantity.

\subsubsection{Planetary Periods}
\label{periods}
\begin{figure}[ht]
   \begin{center}
         \includegraphics[width=3.3in]{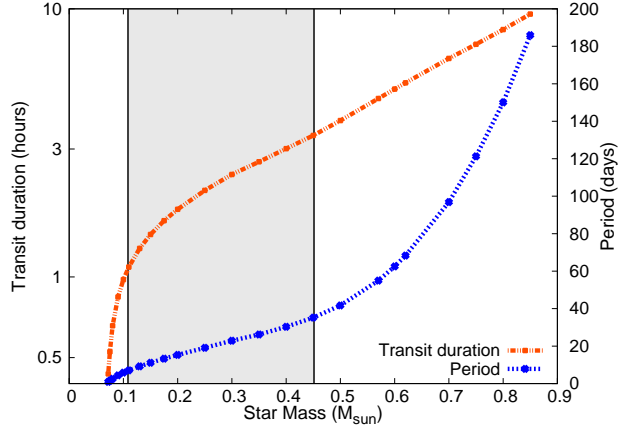}
   \caption{ \footnotesize Transit durations and orbital periods of habitable-zone (HZ) super-Earths for varying masses of M stars. Our focus for the HZ is in the mass range delimited by the grey rectangle: between 0.11 and 0.45 $M_{\odot}$, with orbital periods of 7 to 35 days and are optimal targets, as seen Section \ref{sec:stars}.
For consistency, we use the same stellar types for hot and warm super-Earths. In these cases, the transit and period durations will clearly be shorter (see Tables \ref{tab:se850k} and \ref{tab:se500k} for respective ranges).  }
   \label{fig:star_period}
   \end{center}
   \end{figure}

Figure \ref{fig:star_period} shows periods and transit durations for habitable-zone super-Earths orbiting a range of M stars.
These values clearly depend on orbital distances which are computed by fixing the planet temperature and stellar type. We define the radiation in and out of the planet (in Watts):
\begin{eqnarray}
Rad_{in} &=&\pi R_{pl}^2 L_* f(1-A)\\
Rad_{out} &=&4\pi R_{pl}^2 \sigma T_{pl}^4
\end{eqnarray}
where the planetary albedo is $A=0.3$, and a small greenhouse effect contribution of $\epsilon=0.7$ is assumed. The distance of the habitable-zone for each M star type was estimated by considering an average surface temperature for the planet $T_{pl}=287 K$. More specifically:
we started with the bolometric luminosity of the star at a distance  $a$ (semi-major axis): $L_*=(4 \pi R_*^2 \sigma T_*^4) / (4 \pi a^2)$. 
By equating the radiation in and out of the planet, we obtain an expression for the planetary effective temperature:
\begin{equation}
T_{pl} = T_{*} \left({\sqrt{\frac{\left( 1 - A \right)}{\epsilon}} \frac{R_*}{2 a}}\right)^{\frac{1}{2}}
\end{equation}
by imposing $T_{pl} = 287K$ and assuming different values for the stellar temperature: $2900 K<T_{*}<3900 K$, we rearrange this equation to calculate the semi-major axis $a$ and the planetary period $P = 2\pi \sqrt{(a^3/GM)}$.
For the other cases presented in this paper, period and transit durations were obtained either using similar calculations, or from observations.
For the transit duration $t_t$ we assume a circular, edge-on orbit. From Seager and Ornelas (2002) we obtain: 
\begin{equation}
t_{t} = \frac{P R_*}{\pi a} \sqrt{\left( 1 + \frac{R_{pl}}{R_*} \right)^2 - b^2}
\label{eq:ttransit}
\end{equation}
For the general case the impact parameter $b$  was set to zero, unless otherwise specified,  so that eq. (\ref{eq:ttransit}) simplifies to:
\begin{equation}
t_{t} = \frac{P}{\pi} \left(\frac{R_* + R_{pl}}{a} \right)
\end{equation}
\subsection{Estimating the integration time}
The integration time needed to observe specific targets  depends on: 
\begin{itemize}
\item{the parent star: spectral class, type, magnitude in a specified spectral region}
\item{the contrast between the parent star and the companion planet in the observed spectral interval; this can be estimated from known observed or simulated objects}
\item{the observational requirements: spectral region, resolution and signal to noise ratio}
\item{the telescope characteristics: primary mirror diameter, overall transmission, coverage and sensitivity of the detectors}
\item{the focal plane array characteristics during observation: number of pixels used per spectral resolution element, readout time, quantum efficiency, full well capacity, saturation threshold, dark current, readout noise}
\end{itemize}
We consider then the flux of photons from the planet.
This flux (given in photons/seconds/m$^2$ in the whole spectral interval) is converted into electrons/pixel/seconds/``resolution element'' within the defined spectral region using the following expression
\begin{equation}
F_{e^-} = \frac{F_{\gamma}\cdot A \cdot transmission \cdot QE}{Res \cdot N_{px/Res}}
\end{equation}
where $F_{e^-}$ and $F_{\gamma}$ are respectively the electron and photon fluxes, $A$ is the telescope mirror surface area, $QE$ the quantum efficiency, $Res$ the number of spectral elements in the band (resolution) and $N_{px/Res}$ the number of pixels per resolution element. From here on, $F$ will only refer to the electron flux: $F_{e^-}$. The  \emph{transmission} is the overall fraction of energy that reaches the detector (before conversion to electrons). It includes the telescope and instrument (optical) transmission.
\\
Using these values the time required for one detector  pixel readout is computed:
\begin{equation}
\label{eq:inttime}
t_{ro} = \frac{FWC \cdot saturation}{F_{\star}+F_{pl}+DC}
\end{equation}
where $ro$ stands for read out, $FWC$ for full well capacity, $DC$ for dark current and \emph{saturation} is a fraction of the full well capacity (FWC). Usually, a saturation at 70\% of the FWC is taken into account; that is the limit of electrons that can be accumulated in a single exposure.
\\
The number of readouts required is then computed using the following formula:
\begin{equation}
N_{ro}=(SNR)^2 \cdot \frac{F_{\star}+F_{pl}+DC+(RON^2 / t_{ro})}{F_{pl}^2 \cdot t_{ro} \cdot N_{px/Res}}
\end{equation}\\
where $SNR$ is the signal to noise ratio within the defined spectral band, and $RON$ the detector  readout noise.
For the secondary eclipse case, $F_{pl}$ is the flux emitted or reflected by the planet, while for the primary transit case (explained in section \ref{sec:primary}), $F_{pl}$ corresponds to the amount of flux (written as a negative) absorbed by the planet's atmosphere:
\begin{equation}
F_{pl} = - \frac{\pi{R_{pl}}^2}{\pi{R_{\star}}^2} ((1+\frac{nH^2}{R_{pl}}) - 1) = -\frac{2 n H R_{pl}}{{R_\star}^2}
\end{equation}
where $n$ is an atmospheric absorption factor.\\
With these values, the total integration time is computed by multiplying the duration of a detector  pixel readout by the number of readouts required.
\\
The planet/star flux contrast ratio and the star brightness are the obvious main factors affecting integration times. To estimate the contrast, we have considered observed spectra and simulated synthetic spectra of stellar and planetary atmospheres (Sections \ref{sec:spectra} to \ref{sec:stars}).
\subsubsection{Instrument detector and validation}

Table \ref{tab:instrumentsettings} lists instrument setting values we have assumed for our simulator to cover the four bands in which our results are given in. 

\begin{table*}[h]
\footnotesize
\centering
\begin{tabular}{ l   c  c  c  c }
\hline
Instrument Values & Visible & 2.5 to 5 $\mu$m &  5 - 11 $\mu$m & 11 to 16 $\mu$m\\
\hline
Detector used (SOFRADIR) & CCD & MWIR & LWIR & VLWIR \\
\hline

Full well capacity (electrons) & $2 \cdot 10^6$ & $4 \cdot 10^6$ & $2 \cdot 10^7$ & $5 \cdot 10^6$\\

Dark current (electrons/s/pixel) &  0.1 & 10 & 500 & 300 \\

Quantum efficiency (electrons/photon)&  0.5 & 0.7 & 0.7 & 0.7\\

Readout noise (electrons/pixel/readout) & 10 & 400 & 1000 & 1000\\

Readout time (seconds) & 0.004 & 0.01 & 0.03 & 0.01\\
\hline
Telescope temperature (K) & 0 & 60 & 60 & 60\\

Instrument temperature (K) & 0 & 45 & 45 & 45 \\

Telescope transmission &0.85 & 0.9 & 0.9 & 0.9 \\

Instrument transmission & 0.7 & 0.7 & 0.7 & 0.7\\
\hline
\end{tabular}
\caption{\footnotesize Instrument settings used in our simulations, listed for each observing band used. In addition, the two following settings are the same for all four bands considered: a 30 $\mu$m pixel size and 2 illuminated pixels per spectral element are assumed. For the N band (7.7 to 12.7 $\mu$m) we have used the LWIR setting values.
Note that in the case of the VLWIR detector, we have used a dark current value of 300 electrons/s/pixel considering existing technologies and expected future capabilities. Further discussion on these values can be found in section \ref{sec:instru_discussion}. We give in the appendix two other options, compatible with a 1.2 m telescope, and a different selection of detectors and instrument parameters.}
\label{tab:instrumentsettings}
\end{table*}
For validating our tool, we have incorporated in our instrument simulator the parameters of Hubble NICMOS, and compared our results for hot gaseous planets with observed data from NICMOS.
We obtained results in excellent agreement with the observed data.

\section{Results}
\label{results}
We present our results ordered by planetary temperature: hot, warm and temperate (habitable-zone, H-Z).
For our key examples we have calculated the flux contrast by using synthetic models (see sections \ref{sec:primary} and \ref{sec:contrast}), which either fit existing observations or are extrapolated from our knowledge of the Solar System planets.
For feasibility studies we prefer to adopt cruder estimates of atmospheric contributions (i.e. blackbody curves) rather than detailed simulations of each specific case.
Plots of flux contrasts are given for each case, accompanied by integration times represented as ``number of transits'' (based on transit durations and orbital periods, see section \ref{periods}), with a maximum number of transits indicated. This number is estimated by dividing the nominal lifetime of a mission (we consider 5 years here) by the orbital period for each target. 
For each case, integration times are given over a range of stellar magnitudes.
The signal-to-noise and resolution (SNR/Res) values vary from table to table, from R=300 to R=10, and SNR=50 to SNR=5. For each target, these values were selected to optimise the scientific return across the magnitude range considered. The selected SNR and Resolution values are in most cases dictated by the ``limiting cases'', i.e. the most difficult star+planet combinations to be observed in a specific class of objects. In most tables, the SNR/Res values can be raised for the bright targets, and lowered to curb the integration times for fainter objects. 
The outcome of our study is summarised in the MIR by showing results averaged over the 7.7 to 12.7 $\mu$m spectral window (equivalent to the classical Johnson photometric N-band). In addition, we provide in the appendix results averaged over three spectral bands (5-8.3, 8.3-11, 11-16$\mu$m), the reader may compare performances of various bands for the listed targets. For hot planets, observations in the NIR (2.5 to 5$\mu$m band) become feasible (see section \ref{sec:contrast} with equation \ref{eq:visible}) and planets close to their star can be easily probed in the visible. In such cases, the MIR integration times are followed by NIR and visible results.
\subsection{Hot planets}
\begin{figure}[h]
   \begin{center}
	\plottwo{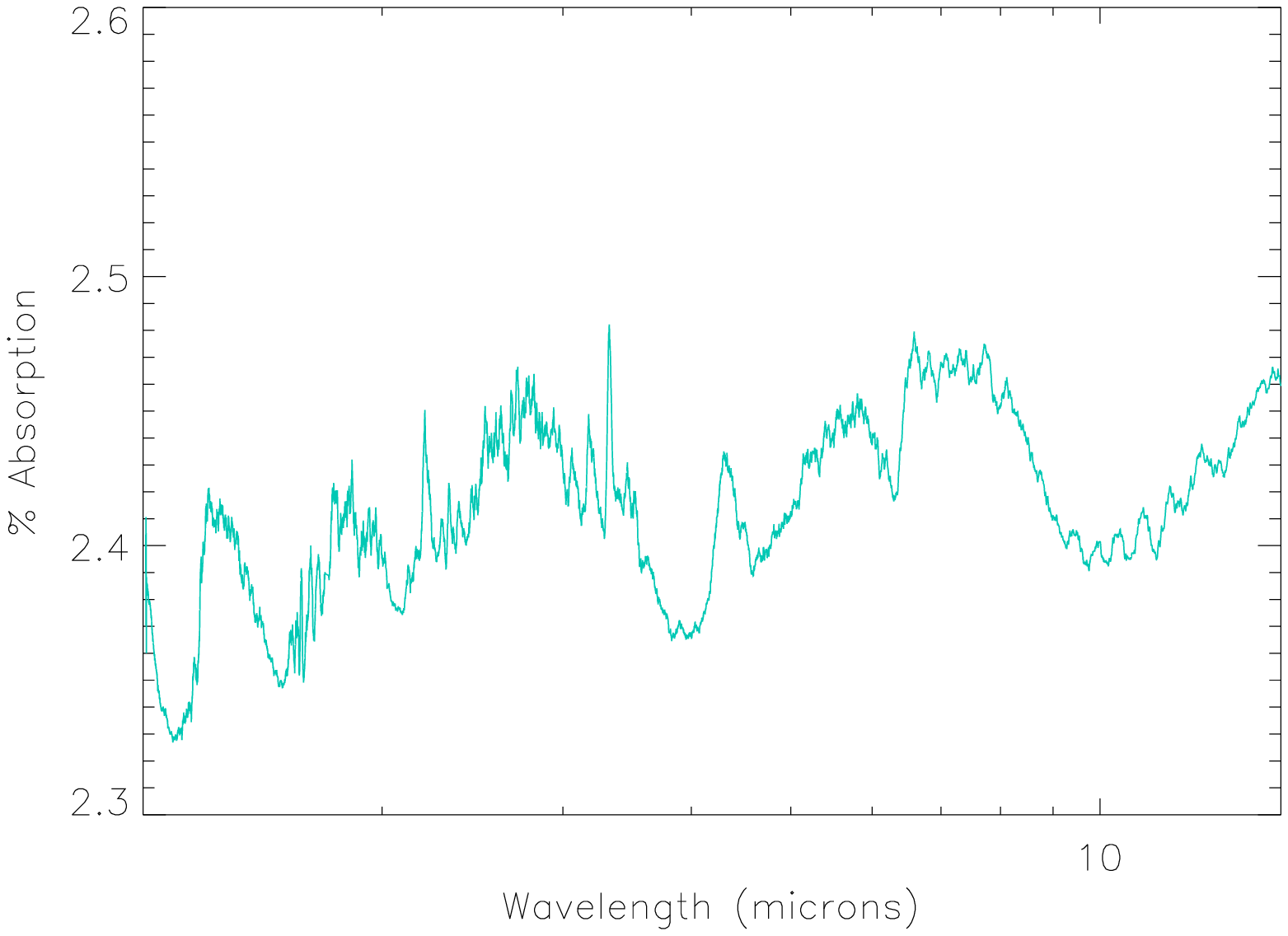}{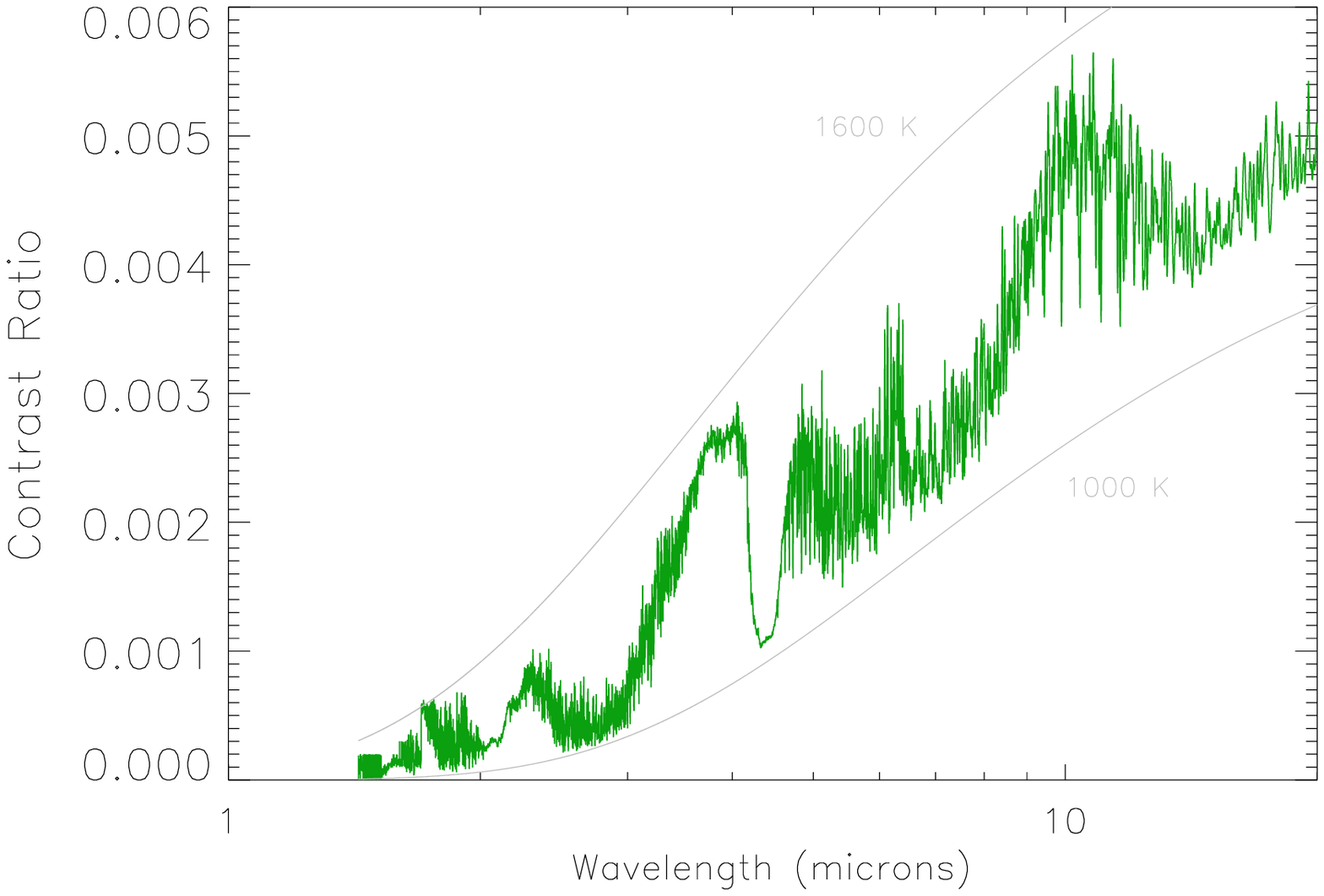}
      \caption{ \footnotesize Modelled transmission and emission spectra of HD 189733b \citep{tinetti_faraday}, a hot-Jupiter around a K1/2V star, mag. V=7.67. \emph{Left:} \% absorption of the stellar flux occulted by the planetary atmosphere during the primary transit (transmission spectrum). \emph{Right:} Contrast ratio of the flux from the planet (emission spectrum) over the flux from the star. Blackbody curves at 1000 K and 1600 K are plotted in grey.}
      \label{fig:hd189}
   \end{center}
   \end{figure}

 \paragraph{Gas giants:}  as a template for the hot Jupiter case, the observed hot gas giant HD 189733b is used. 
A modelled transmission spectrum analogue of primary transit observations and a planet/star contrast ratio, analogue of secondary eclipse measurements, are considered for our simulations (Fig. \ref{fig:hd189}). For both cases, integration times are listed in units of number of transits in Table \ref{tab:189primary}, where the modelled hot Jupiter is presented orbiting a sample of stars: a Sun-like G2V star, a warmer F3V star and HD 189733, a K1/2V type star \citep{bouchy189}. HD 189733 has a magnitude in V of 7.67. We extrapolate our results from mag V=5 to V=9, with a resolving power of R=300 and a signal-to-noise ratio SNR=50, chosen for the secondary eclipse, and R=100 and SNR=50 for the primary transit.
\newline
\begin{table}[ht]
\footnotesize
\centering 
\begin{tabular}{ | l c c | c | c  c c|ccccc| }
   \multicolumn{12}{l}{Hot-Jupiters --Secondary eclipse, R=300, SNR=50, MIR } \\
  \hline
Star & T & R &Contrast  & Period & $\tau_{transit}$ & Max. n* & \multicolumn{5}{ c|}{Integration time (n. transits)} \\
type & (K)&($R_{\odot}$) &($*10^{-3}$) & (days) & (hours) & (transits) &   V=5 & V=6 & V=7 & V=8 & V=9 \\
\hline
F3V$^\dagger$ & 6740  & 1.56 & 1 & 8.4 & 2.9 &     218  &  7  & 18  & 51 & 156 & \textit{lower R} \\
\cline{1-3} \cline{8-12}
G2V & 5800 & 1 & 2.9 & 3.2 & 2.36 &         570 & 0.7 & 1.8 & 4.7 & 14 & 45 \\
\cline{1-3} \cline{8-12}
K1V$^\dagger$ & 4980 & 0.8 & 5.6 & 2.21 & 1.8 &         826    &  0.2 & 0.4 & 1 & 2.9 & 9 \\
\hline
\multicolumn{11}{l}{Hot-Jupiters --Primary transit, R=100, SNR=50, MIR} \\
\hline
F3V &  6740 & 1.56 & 0.28 & 8.4 & 2.9 &     218     & 32 & 82 & 213 & \multicolumn{2}{c|}{\textit{lower R}} \\
\cline{1-3} \cline{8-12}
G2V & 5800 & 1 & 0.68 & 3.2 & 2.36 &         570     & 4 & 10 & 26 & 70 & 198\\
\cline{1-3} \cline{8-12}
K1V &  4980 & 0.8 & 1 & 2.21 &  1.8 &    826     & 1.6 & 4 & 10 & 26 & 72 \\
\hline
   \multicolumn{12}{l}{Hot-Jupiter in NIR -- Secondary eclipse, R=300, SNR=50, NIR } \\
\hline
K1V & 4980 & 0.8 & 2.6 & 2.21 & 1.8 & 		826	&  0.1 & 0.2 & 0.6 & 1.4 & 3.5 \\
\hline
\end{tabular} 
\caption{ \footnotesize Integration times (in units of ``number of transits") needed to obtain the specified SNR and spectral resolution for a given stellar type/brightness (in Mag. V). The upper table lists results for the secondary eclipse scenario in the MIR (equivalent to the classical Johnson photometric N-band) followed by primary transit results in the MIR, and secondary eclipse results in the NIR (between 2.5 and 5 $\mu m$). 
$\tau_{transit}$ is the transit duration given in hours, and ``\textit{lower R}" stands for target observable at lower resolution. $\dagger$: Planet/star systems marked by this sign have additional results listed in the appendix. $*$: The maximum number of transits is computed by dividing a plausible mission lifetime (5 years assumed) by the duration of the planet orbital period.}
\label{tab:189primary} 
\end{table}
 \paragraph{Neptunes:} Neptune-like planets are expected to have a similar atmospheric composition to the gas-giants with a smaller radius ($R\sim$0.35 $R_{j}$).
While we do not directly present results for these targets, by comparison with the hot Jupiter scenario, integration times will be typically similar in the primary transit scenario and higher in the secondary eclipse scenario given the relatively smaller radius of the planet.

\begin{figure}[ht]
\begin{center}
	\plottwo{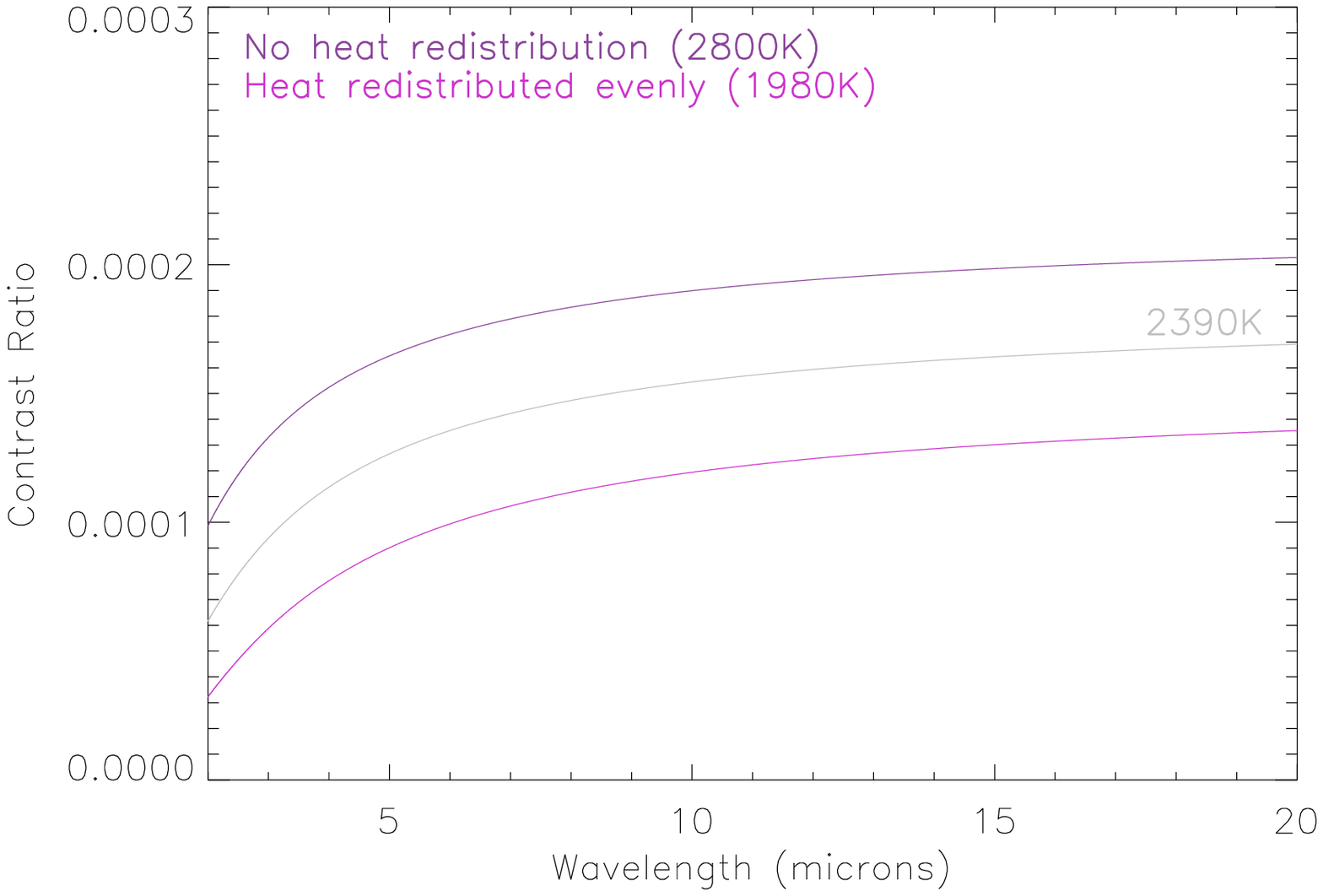}{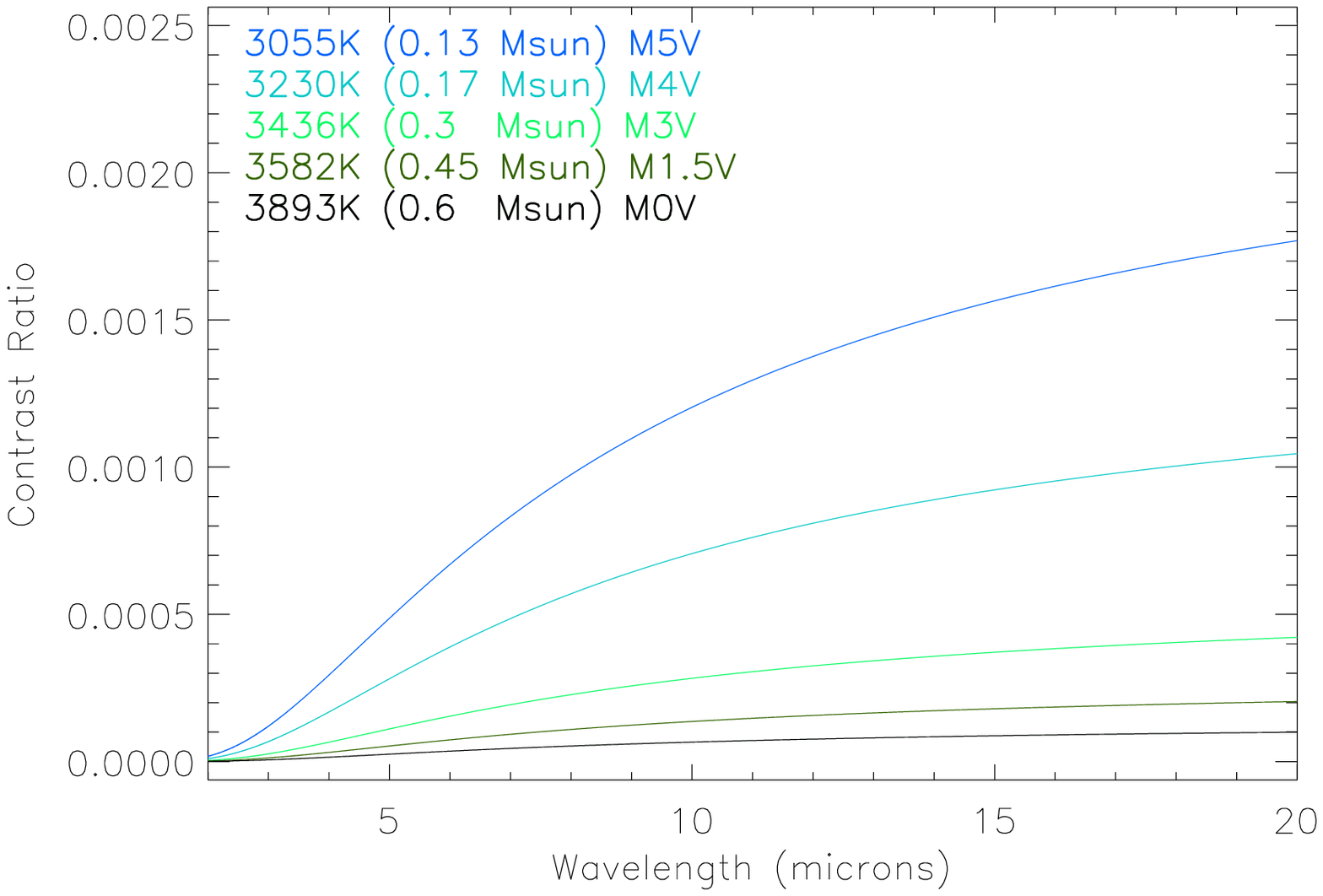}
      \caption{ \footnotesize \emph{Left:} Secondary eclipse simulated signal for 55 Cancri e, a 2.1 $R_\oplus$ hot super-Earth orbiting a G8V star. The atmospheric temperature could vary between 2800 K and 1980 K, depending on the heat redistribution \citep{winn2011}. Both possibilities are presented, alongside an intermediate case of a 2390 K atmosphere used for our results.  \emph{Right:} Secondary eclipse signal for a hot super Earth (850K, 1.6 $R_\oplus$) orbiting a selection of M stars (from M1.5V to M5V). For the two figures, both the planet and the stellar contributions here are estimated as black-bodies. While this description is too simplistic to capture the properties of a real, specific case, for feasibility tests we do not want to rely on too narrow assumptions. }
      \label{fig:se850k}
   \end{center}
\end{figure}

 \paragraph{Super-Earths:} we show here two examples: a 2.1 $R_{\oplus}$ very hot planet in orbit around a G8V star, 55 Cancri e \citep{winn2011}, and a 1.6 $R_{\oplus}$, 850 K planet in orbit around a range of M stars with temperature varying between $3055 \leq T \leq 3582 K$. For the latter case, we approximated the planet/star fluxes with black-body curves to assess feasibility. As mentioned in section \ref{sec:primary}, primary transit observations for a planet with high gravitational pull might be out of reach (55 Cancri \nolinebreak e is reported to be $\sim 8.5 M_\oplus$), for this reason we focus on secondary eclipses only.
Planet to star flux contrasts are plotted in Figure \ref{fig:se850k} (55 Cancri \nolinebreak e left, 850 K super-Earth right), accompanied by integration times in Table \ref{tab:se850k} in the MIR and NIR. For both bands a resolution of R=40 and SNR=10 were selected.
\newline
\begin{table}[ht]
\centering
\footnotesize
\begin{tabular}{| l  c c |c | c   c  c | c c c c c |}
\multicolumn{12}{l}{Hot super-Earths --Secondary eclipse. R=40, SNR=10, MIR}\\
\hline
Star & T & R &Contrast & Period & $\tau_{transit}$ & Max. n* & \multicolumn{5}{ c|}{Integration time (n. transits)} \\
type & (K)&($R_{\odot}$) & ($*10^{-4}$) & (days) & (hours) & (transits) &  K=5 & K=6 & K=7 & K=8 & K=9\\
\hline
M0V & 3893 &0.57 & 0.7 & 2.17 & 1.6 & 840 &  38 & 97 & 253 & 689 & \textit{lower R.} \\
\cline{1-3} \cline{8-12}
M1.5V$^\dagger$ & 3582 &0.42 & 1.4 & 1.22 & 1.1 & 1494 &  13 & 33 & 87 & 236  & 707 \\
\cline{1-3} \cline{8-12}
M3V & 3436  &0.30 & 2.9 & 0.79 & 0.8 	& 2300 &  4.2 & 11 & 28 & 75 & 225 \\
\cline{1-3} \cline{8-12}
M4V & 3230 &0.20 & 7.2 & 0.46 & 0.5   & 3955 &  1 & 2.7 & 7 & 19 & 58 \\
\cline{1-3} \cline{8-12}
M5V$^\dagger$ & 3055 &0.16 &12.2 & 0.25 & 0.4  	& 7450 &  0.5 & 1.2 & 3.1 & 8 & 25 \\
\hline
\multicolumn{12}{l}{Hot super-Earths in NIR --Secondary eclipse, R=40, SNR=10, NIR}\\
\hline
M0V & 3893 & 0.57 & 0.1 & 2.17 & 1.6 & 840 & 199 & 499 & \multicolumn{2}{ c}{\textit{lower R.}} &  \\
\cline{1-3} \cline{8-12}
M1.5V & 3582 &0.42 & 0.3 & 1.22 & 1.1 & 1494 &  32 & 81 & 203 & 509 & 1279 \\
\cline{1-3} \cline{8-12}
M3V & 3436  &0.30 & 0.5 & 0.79 & 0.8 	& 2300 &  15 & 39 & 97 &  243 & 611 \\
\cline{1-3} \cline{8-12}
M4V & 3230 &0.20 & 1.4 & 0.46 & 0.5   & 3955 &  3 & 8 & 19 & 18 & 121 \\
\cline{1-3} \cline{8-12}
M5V & 3055 &0.16 & 2.5 & 0.25 & 0.4  	& 7450 &  1.1 & 2.8 & 7 & 18 & 45 \\
\hline
\multicolumn{12}{l}{Hot super-Earth --example of 55 Cancri e in secondary transit, R=40, SNR=10, MIR}\\
\hline
Star & T & R &Contrast & Period & $\tau_{transit}$ & Max. n* & \multicolumn{5}{ c|}{Integration time (n. transits)} \\
type & (K)&($R_{\odot}$) & ($*10^{-4}$) & (days) & (hours) & (transits) &  V=5 & V=6 & V=7 & V=8 & V=9\\
\hline
G8V & 5243 & 0.95 & 1.6 & 0.74 & 1.76  	& 2467 & 1.4  & 3.4 & 9 & 22 & 58 \\
\hline
\end{tabular}
\caption{ \footnotesize Integration times (in units of ``number of transits") needed to obtain the specified SNR and spectral resolution for a given stellar type/brightness (in Mag. K when orbiting M dwarfs, Mag. V when orbiting G star). The upper table lists results for the secondary eclipse scenario in the MIR, followed by secondary eclipse results in the NIR.
$\tau_{transit}$ is the transit duration given in hours, and  ``\textit{lower R}" stands for target observable at lower resolution. \emph{$\dagger,*$: See Table \ref{tab:189primary} caption.}
 }
\label{tab:se850k}
\end{table}
\paragraph{Observations in the visible:} 
we present here two cases: the case of a hot Jupiter and the case of a hot super-Earth. The reasons for our choice are based on Eq. \ref{eq:visible}: reflected light is more prominent for planets close to their star.
For the case of the hot super-Earth, we selected a 1.6 $R_\oplus$ planet with a fixed temperature of 850 K and varying albedo values. For the case of the hot Jupiter, we present a fixed orbital distance with varying albedo values (corresponding to temperatures $\sim1200 - 1500$ K).
Notice that the emission from the planet is negligible at these temperatures when compared with reflection in the visible.
Results are given in Tables \ref{tab:VISint} and \ref{tab:VISint2}, with R=40 and SNR=20 for the hot Jupiter, and R=20 and SNR=10 for the hot super-Earth.
\begin{table}[ht]
\footnotesize
\centering
\begin{tabular}{| c  | c | c | c c c | c c c c c |}
\multicolumn{7}{l}{Visible band hot Jupiter. With $\zeta=1$, R=40, SNR=20}\\
\hline
Albedo & $a$ & Contrast & Period & $\tau_{transit}$ & Max. n* &  \multicolumn{5}{ c|}{Integration time (n. of transits)} \\
Value & (a.u.) & ($*10^{-4}$) & (days) & (hours) & (transits) & V=5 & V=6 & V=7 & V=8 & V=9\\
\hline
0.1 & \multirow{4}{*}{0.031} & 0.31 & \multirow{4}{*}{4.7} &\multirow{4}{*}{2.36} & \multirow{4}{*}{570} &   7 & 18 & 44 & 110 & 278 \\
\cline{1-1}\cline{3-3}\cline{7-11}
0.3 &  & 0.92 & &&&  0.8 & 1.9 & 4.9 & 12 & 31 \\
\cline{1-1}\cline{3-3}\cline{7-11}
0.5 &  & 1.54 & &&&  0.3 & 0.7 & 1.8 & 4.5 & 11  \\
\cline{1-1}\cline{3-3}\cline{7-11}
0.7 &  & 2.16 & &&&  0.1 & 0.4 & 0.9 & 2.3 & 6 \\
\hline
\end{tabular}
\caption{ \footnotesize Integration times (in units of ``number of transits'') for a hot Jupiter observed in the visible around a G2V star. The orbital distance is fixed and the planetary temperature varies with the albedo. For the studies presented here, we have considered full illumination ($\zeta=1$), and values of R=40 and SNR=20. When the planet is not fully illuminated ($\zeta<1$), longer integration times are needed for the same parameters. \emph{$*$: See Table \ref{tab:189primary} caption.}}
\label{tab:VISint}
\end{table}
\begin{table*}[ht]
\footnotesize
\centering
\begin{tabular}{| c  | c | c | c c c | c c c c c |}
\multicolumn{11}{l}{Visible band hot super-Earth, with $\zeta=1$, R=20, SNR=10}\\
\hline
Albedo & $a$ & Contrast & Period & $\tau_{transit}$ & Max. n* &  \multicolumn{5}{ c|}{Integration time (n. of transits)} \\
Value & (a.u.) & ($*10^{-4}$) & (days) & (hours) & (transits) & K=5 & K=6 & K=7 & K=8 & K=9\\
\hline
0.1 & 0.006 & 0.15 & 0.47 & 0.8 & 3916 &  426 & 1161 & 2917 &\multicolumn{2}{c |}{\textit{lower R.}} \\
\hline
0.3 &  0.006 & 0.58 & 0.39 & 0.7 & 4728 &  35 & 89 & 223 & 560 & 1407 \\
\hline
0.5 & 0.005  & 1.35 & 0.30 & 0.7 & 6085 &  7 & 16 & 41 & 103 & 260  \\
\hline
0.7 &  0.004 & 3.16 & 0.20 & 0.6 & 8927 &  1.4 & 3.5 & 9 & 22 & 55 \\
\hline
\end{tabular}
\caption{ \footnotesize Integration times (in units of ``number of transits'') for a hot super-Earth (850 K) observed in the visible around a M4.5V star. Here the planetary temperature is fixed and the orbital distance varies with the albedo. For the studies presented here, we have considered full illumination ($\zeta=1$), and values of R=20 and SNR=10. When the planet is not fully illuminated ($\zeta<1$), longer integration times are needed for the same parameters. ``\textit{lower R}" stands for target observable at lower resolution. \emph{$*$: See Table \ref{tab:189primary} caption.} }
\label{tab:VISint2}
\end{table*}
\subsection{Warm planets}
\begin{figure}[h]
   \begin{center}
   \plottwo{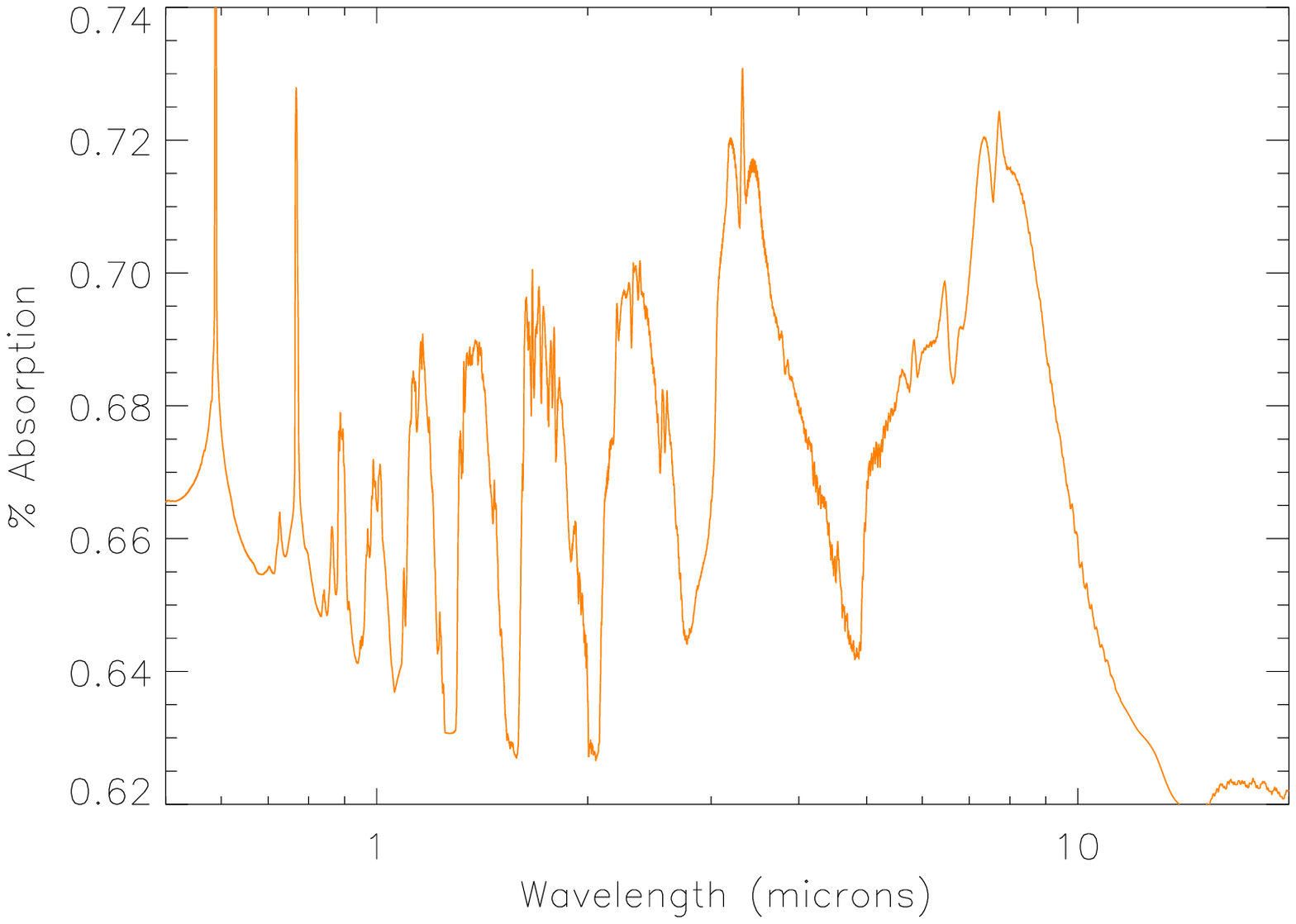}{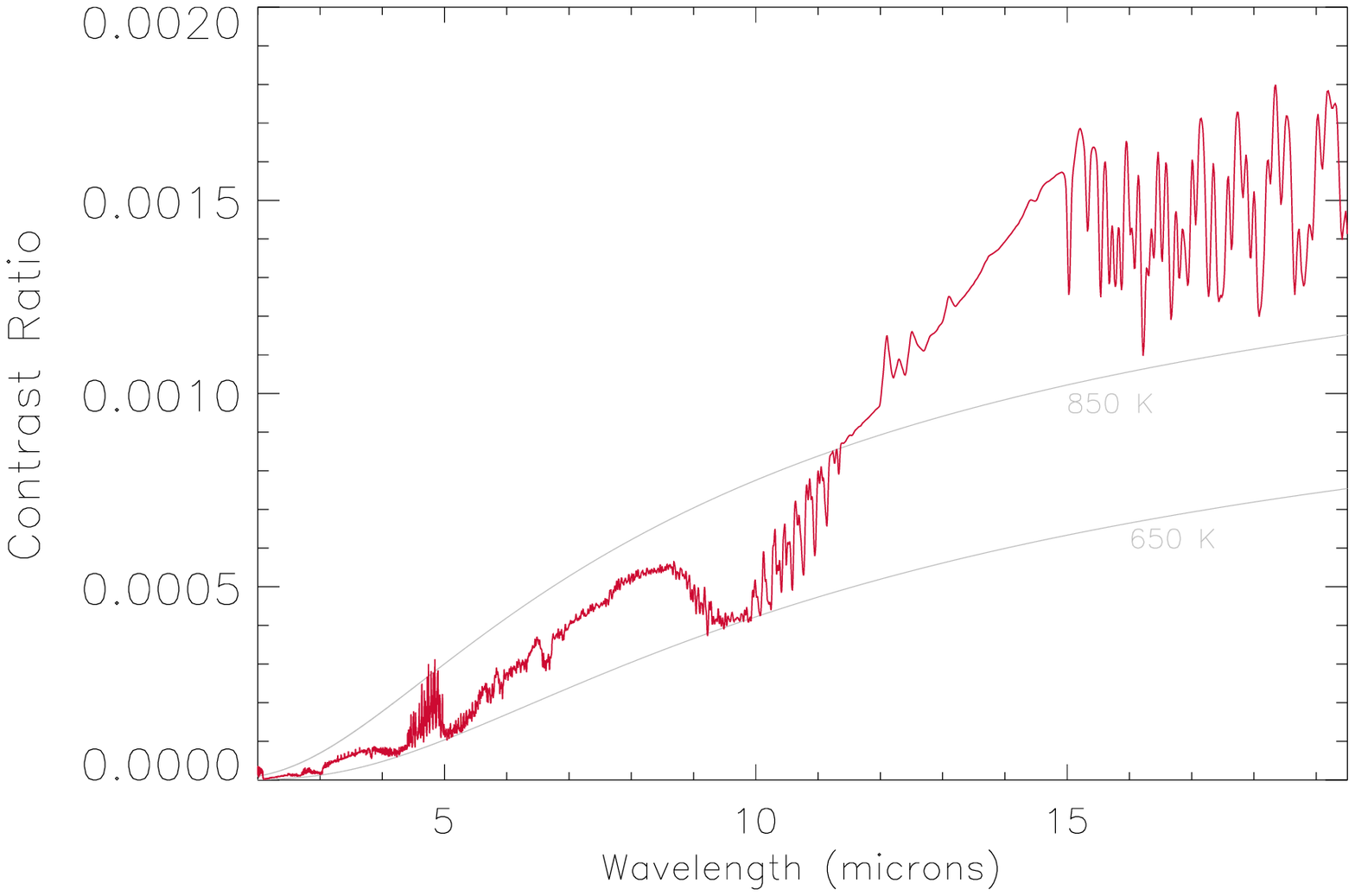}
      \caption{ \footnotesize Modelled GJ 436b \citep{beaulieu2011,stevenson2010}, a warm Neptune around a M2.5V star, mag. K=6.07: \emph{Left:} \% absorption of the stellar flux occulted by the planetary atmosphere during the primary transit. \emph{Right:} Contrast ratio of the flux from the planet over the flux from the star. Blackbody curves at 650 K and 850 K are plotted in grey.}
   \label{fig:gl436primary}
   \end{center}
   \end{figure}

\paragraph{Gas giants:} In this section we focus on Neptunes and super-Earths, skipping warm gas giants, which fall between the categories of hot Jupiters and warm Neptunes.
\paragraph{Neptunes:} we considered as example of a warm Neptune GJ 436b, a 4 $R_\oplus$ planet around a M2.5V dwarf star, with a radius of 0.46 $R_\odot$ and magnitude in K of 6.07  \citep{butler436,gillon2007}.
Spitzer photometric data have been analysed and interpreted (by \citet{beaulieu2011,stevenson2010,knutson2011}), observed results captured by simulated spectra are shown in Figure \ref{fig:gl436primary} (primary transit left, secondary eclipse right). Integration times for a primary transit and secondary eclipse of such a warm Neptune-like planet follow in Table \ref{tab:436primary}.
\newline
\begin{table}[h]
\centering
\footnotesize   
\begin{tabular}{| lcc|c|ccc|c|ccccc | }
\multicolumn{13}{l}{Warm Neptune --Secondary eclipse, R=50-100, SNR=30-50, MIR}\\
\tableline
Star & T & R & Contrast  & Period & $\tau_{trans}$ & Max. n & R/SNR &\multicolumn{5}{ c|}{Integration time (n. transits)} \\
type & (K)&($R_{\odot}$) & $*10^{-4}$ & (days) & (hrs) & (trans.) &  & K=5 & K=6 & K=7 & K=8 & K=9\\
\hline
\multirow{2}{*}{M2.5V$^\dagger$} & \multirow{2}{*}{3684} & \multirow{2}{*}{0.46} &  \multirow{2}{*}{4.6}& \multirow{2}{*}{2.64} & \multirow{2}{*}{1.03} & \multirow{2}{*}{691}  & 100/50 & 80 & 207 & 563 & \multicolumn{2}{c|}{\emph{lower Res.}}   \\
\cline{8-13}
 &  & &  &  & &   &50/30 & 14 & 36 & 95 & 263 & \emph{low} R  \\
\hline
\multicolumn{11}{l}{Warm Neptune --Primary transit, R=50-100, SNR=30-50, MIR}\\
\hline
\multirow{2}{*}{M2.5V} & \multirow{2}{*}{3684} & \multirow{2}{*}{0.46} &  \multirow{2}{*}{10}& \multirow{2}{*}{2.64} & \multirow{2}{*}{1.03} & \multirow{2}{*}{691}  & 100/50 & 17 & 44 & 120 & 358  & \emph{low} R \\
\cline{8-13}
&  &  &  &  &  & & 50/30 & 3 & 8 & 20 & 56 & 173 \\
\hline
\end{tabular}
\caption{ \footnotesize Integration times (in units of ``number of transits") needed to obtain the specified SNR and spectral resolution for a given stellar type/brightness (in Mag. K). The upper table lists results for the secondary eclipse scenario in the MIR, followed by primary transit results in the MIR. Both tables show two selections of SNR and resolution values.
$\tau_{transit}$ is the transit duration given in hours, and  ``\textit{lower R}" stands for target observable at lower resolution. \emph{$\dagger,*$: See Table \ref{tab:189primary} caption.}} 
\label{tab:436primary}
\end{table}
\newline
   \begin{figure}[h]
   \begin{center}
   \plottwo{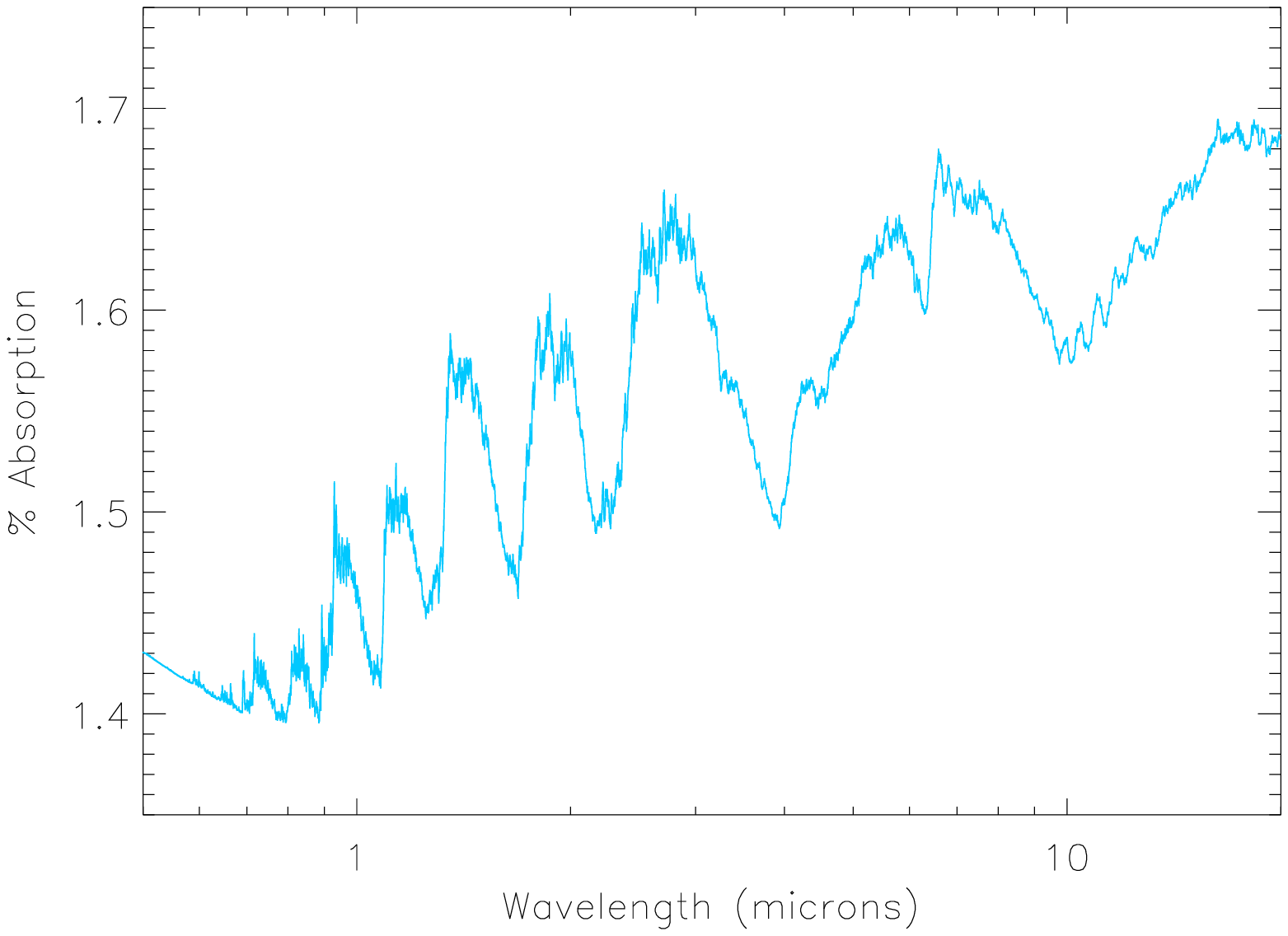}{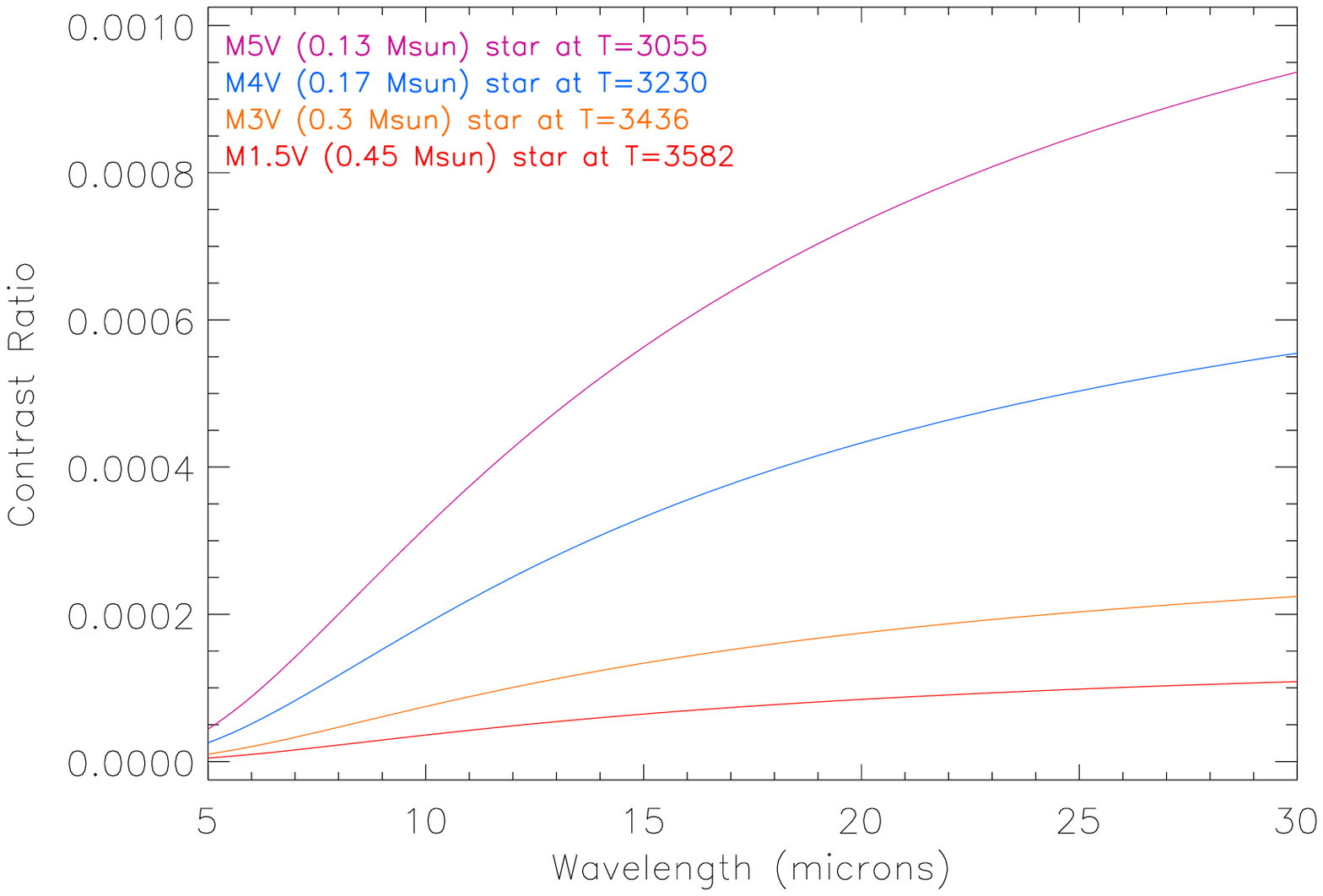}
   \caption{ \footnotesize \emph{Left:} simulated transmission spectrum for the warm super-Earth GJ 1214b, in units of \% absorption of the stellar flux. \emph{Right:} secondary eclipse signal from a warm Super Earth (500 K, 1.6 $R_\oplus$) orbiting a range of M stars, from M1.5V to M5V with temperatures ranging from 3055 K to 3582 K.}
   \label{fig:se500k}
   \end{center}
   \end{figure} 
\paragraph{Super-Earths:} GJ 1214b \citep{charbonneau_super-earth_2009} is a perfect example for the case of a warm super-Earth orbiting a M star. We show in Fig. \ref{fig:se500k} a simulated transmission spectrum of this planet. Since the available observations for this specific planet are not enough to constrain its true composition and atmospheric characteristics \citep{bean2010}, our simulations here just show a possible scenario.
We also present in Fig. \ref{fig:se500k} planet/star flux contrasts for a 1.6 $R_\oplus$, 500 K planet in orbit of a range of M stars (from M1.5V to M5V with temperatures ranging from 3055 K to 3582 K). Both the planet and the stellar contributions here are estimated as blackbodies, and only secondary eclipse results are presented. The integration times are listed in Table \ref{tab:se500k} in the MIR, with R=40 and SNR=10.\\
     \begin{table}[h]
   \centering
   \footnotesize
   \begin{tabular}{ | l cc | c | ccc | c c c c c | }
    \multicolumn{12}{l}{Warm super-Earths --Secondary eclipse, R=40, SNR=10, MIR}\\
    \hline
Star & T & R & Contrast & Period & $\tau_{transit}$ & Max. n* & \multicolumn{5}{ c |}{Integration time (n. transits)} \\
type & (K)&($R_{\odot}$) &$*10^{-4}$ & (days) & (hours) & (transits) &   K=5 & K=6 & K=7 & K=8 & K=9\\
\hline
M1.5V & 3582 &0.42 & 0.4 & 6 & 1.9 &	304	 & 52 & 131 & 335 & \textit{low R}&\textit{ph}\\
\cline{1-3} \cline{8-12}
M3V & 3436  &0.30 & 0.8 & 3.9 & 1.3 & 468	 & 18 & 44 & 114 & 298 & \textit{low R}  \\
\cline{1-3} \cline{8-12}
M4V$^\dagger$ & 3230 &0.20  & 1.9 & 2.27 & 0.9 & 804	 &  4 & 10 & 26 & 69 & 192 \\
\cline{1-3} \cline{8-12}
M5V$^\dagger$ & 3055 &0.16 & 3.3  & 1.57 & 0.7 & 1163	 & 1.8 & 4.6 & 12 & 31 & 85 \\
\hline
    \multicolumn{12}{l}{Warm super-Earth --example of GJ1214b in primary transit, R=40, SNR=10, MIR}\\
    \hline
M4.5V & 2949 &0.21 & 27  & 1.58 & 0.88 & 1155	 &  0.1 & 0.1  & 0.3  & 0.8  & 2.3  \\
\hline
\end{tabular}
  \caption{ \footnotesize Integration times (in units of ``number of transits") needed to obtain the specified SNR and spectral resolution for a given stellar type/brightness (in Mag. K). The upper table lists results for the secondary eclipse scenario in the MIR, followed by secondary eclipse results in the NIR.
  $\tau_{transit}$ is the transit duration given in hours,  ``\textit{lower R}" stands for target observable at lower resolution, and \textit{ph} stands for photometry. \emph{$\dagger,*$: See Table \ref{tab:189primary} caption.}}
     \label{tab:se500k}
\end{table}   

\subsection{Habitable Zone Planets}
\label{sec:hzplanets}
\begin{figure}[h]
   \begin{center}
   \includegraphics[width=3.3in]{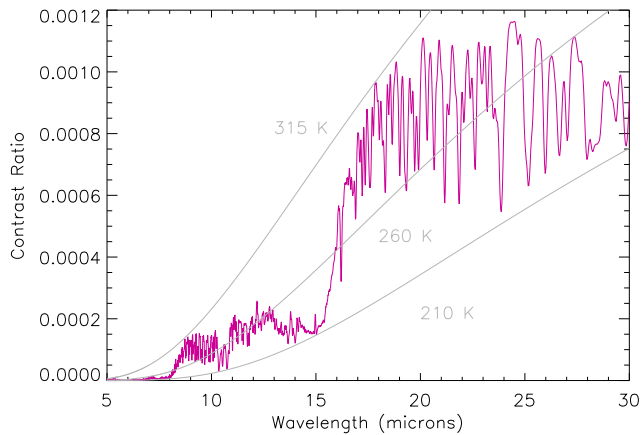}
   \caption{ \footnotesize Secondary eclipse signal from a conceivable habitable-zone Jupiter around a K4V, 4780 K star --such as HAT-P-11. 
    Blackbody curves at 210 K, 260 K and 315 K are plotted in grey.}
   \label{fig:warmjupiter}
   \end{center}
   \end{figure}

\begin{figure}[h]
   \begin{center}
   \includegraphics[width=3.1in]{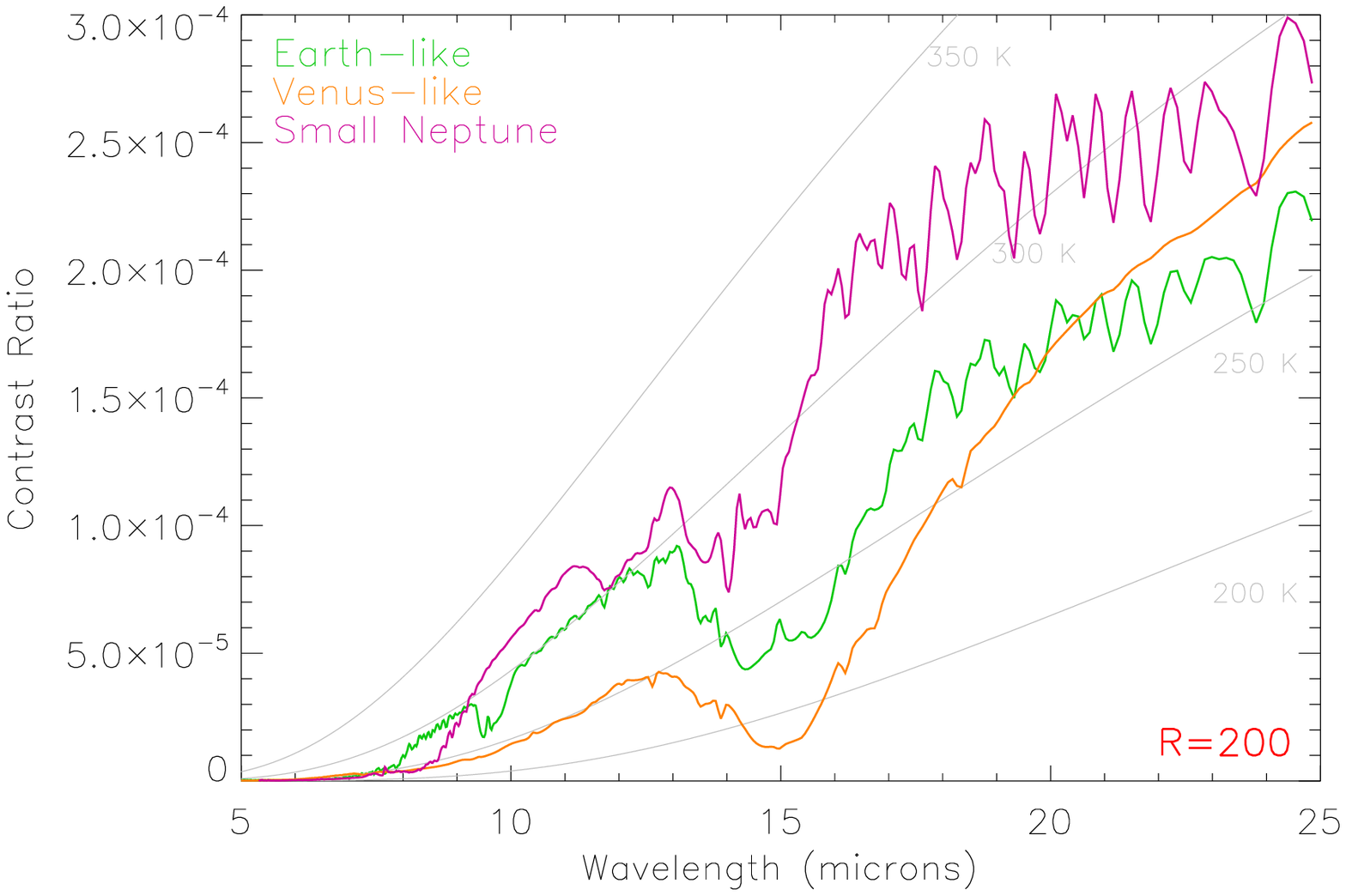}
   \includegraphics[width=3.1in]{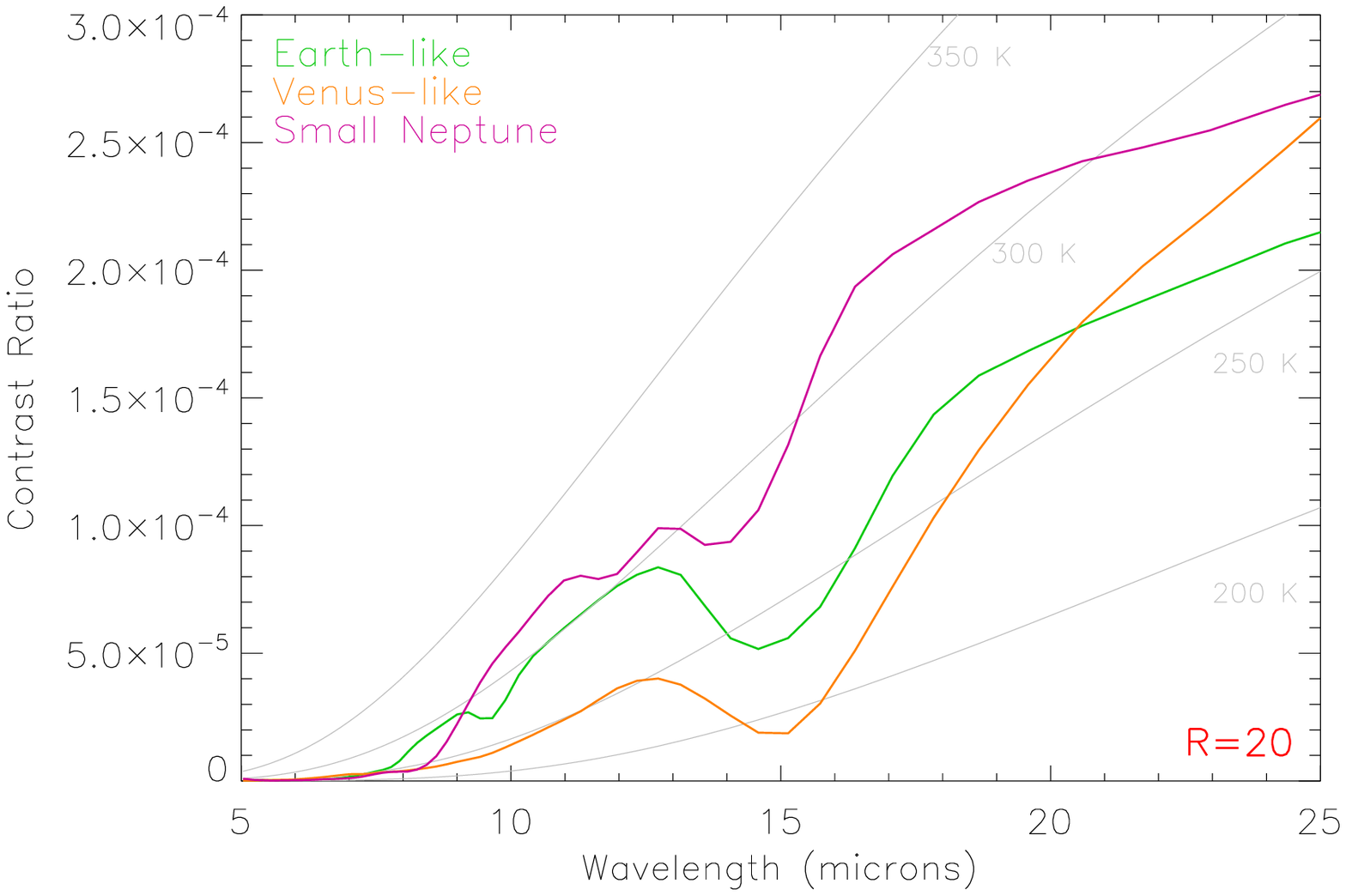}
   \caption{ \footnotesize \emph{Left:} Earth-like, Venus-like and small Nepture secondary eclipse spectra at R=200, with marked blackbody contrast curves as temperature indicators (from left to right: 350, 300, 250 and 200 K). The three atmospheres belong to a 1.8 $R_\oplus$ super Earth around an M4.5V star (at T=3150 K). \emph{Right:} Same case at a resolution of R=20.}
   \label{fig:four}
   \end{center}
   \end{figure}

\paragraph{Gas giants:}we present here the case of a hypothetical ``cool'' Jupiter, in the Habitable-Zone (HZ) of a K4V star. Figure \ref{fig:warmjupiter} shows our simulated secondary eclipse spectrum, with an atmosphere in which we have included water vapour, methane, hydrocarbons, CO and CO$_2$ and a thermal profile with temperature decreasing with altitude. In Figure \ref{fig:warmjupiter}, the departure from the (315 K) blackbody is noticeable. While our assumptions here are reasonable, this is just one possible scenario, and
completeness is beyond the scope of this paper. 
Integration times are listed in Table \ref{tab:hzjup}, for different stellar brightness.  
\newline
      \begin{table}[h]
\centering   
\footnotesize
\begin{tabular}{ | lcc | c | c c c| c | c c c c c| }
\multicolumn{12}{l}{Cool Jupiter --Secondary eclipse, R=20-40, SNR=10, MIR}\\
\hline
Star & T & R & Contrast  &Period & $\tau_{trans.}$ & Max. n* & R/SNR & \multicolumn{5}{ c|}{Integration time (n. of transits)} \\
type & (K)&($R_{\odot}$) & ($*10^{-4}$) & (days) & (hrs) & (transits) & & V=5 & V=6 & V=7 & V=8 & V=9\\
\hline
\multirow{2}{*}{K4V$^\dagger$} & \multirow{2}{*}{4780} & \multirow{2}{*}{0.75} & \multirow{2}{*}{1.5} & \multirow{2}{*}{101.6} & \multirow{2}{*}{6.9} & \multirow{2}{*}{18}  & 40/10 & 0.3 & 0.6 & 1.6 & 4.1 & 11\\
\cline{8-13}
 &  &  &  & &  &  & 20/10 & 0.1 & 0.3 & 0.8 & 2 & 5\\
\hline
\end{tabular}
\caption{ \footnotesize Integration times (in units of ``number of transits") needed to obtain the specified SNR and spectral resolution for a given brightness (in Mag. V). The results are given in the MIR with two selections of SNR and resolution.
 $\tau_{trans.}$ is the transit duration given in hours. Notice that the orbital period for a planet in the HZ of a K4V star is more than 100 days, so the observation can be repeated less than 20 times in 5 years. \emph{$\dagger,*$: See Table \ref{tab:189primary} caption.}} 
\label{tab:hzjup}
\end{table}
\paragraph{Neptunes:} we skip the case of a habitable-zone Neptune, as the secondary eclipse falls between the categories of a HZ Jupiter and a HZ super-Earth. In the case of primary transits, on the contrary, we expect a much more favourable result, as indicated in Table \ref{tab:primaryvssecondary}.
\paragraph{Super-Earths:}here we present a 1.8 R$_{\oplus}$ telluric planet, with three plausible atmospheres, as explained in section \ref{sec:spectra}: Earth-like, Venus-like and hydrogen-rich (i.e. small Neptune).
Figure \ref{fig:four} shows the planet to star flux contrast obtained for a 1.8 R$_{\oplus}$ super-Earth orbiting a M4.5V star with T=3150 K, with the three mentioned atmospheres in two spectral resolutions: R=200 and R=20. Blackbody curves at 200, 250, 300, 350 K are included. 
 The change in contrast for the different atmospheric cases is noticeable: for instance, the presence of water vapour in the Earth-like and small Neptune cases marks a sharper departure from the blackbody curve. H$_2$O, CO$_2$ and ozone absorption are still detectable even at very low resolution, but less abundant hydrocarbon species become more difficult to capture.
 Table \ref{tab:mainresults320} lists integration times in the MIR for the case of a 300 K atmosphere and a range of stars spanning in type and brightness. While a resolution of R=10 and SNR=5 were selected to cover the broadest range of stellar types in the table, the cooler stars in the table will allow for higher SNR/Resolution values.
  \begin{table}[h]
   \centering
   \footnotesize
   \begin{tabular}{ | lcc | c | c  c  c  | c c  c c c |}
 \multicolumn{12}{l}{Habitable Zone super-Earth --Secondary eclipse, R=10, SNR=5, MIR}\\
\hline
Star & T & R & Contrast & Period & $\tau_{transit}$ & Max. n* &  \multicolumn{5}{ c | }{Integration time (in N. transits)} \\
type & (K)&($R_{\odot}$) & ($*10^{-5}$) & (days) & (hours) & (transits) &  K=5 & K=6 & K=7 & K=8 & K=9 \\
\hline
M2.5V	& 3475	& 0.34 & 1.1 &	23.7	& 2.6	& 77		& 54 	& & \multicolumn{3}{c|}{\textit{photometry}}  \\
\cline{1-3} \cline{8-12}
\multirow{2}{*}{M3V} & 3436  & 0.30 & 1.4 &	20.6		& 2.3 	& 88	 	& 37 	& & \multicolumn{3}{c|}{\textit{photometry}} \\
 	& 3380	& 0.26 & 1.9 &	17.3	& 2.1 	&  105		& 22 	& 55 	& \multicolumn{3}{c|}{\textit{photometry}} \\
\cline{1-3} \cline{8-12}
\multirow{2}{*}{M4V$^\dagger$} & 3230 	& 0.20 & 3.5 &	12	& 1.6 	& 152	 	& 9		& 22 	& 54   & 140 & \textit{ph.}\\
 	& 3150 	& 0.17 & 4.6 &	10	& 1.4 	& 182	 	& 6  		& 14 	& 36 & 94 & \textit{ph.} \\
\cline{1-3} \cline{8-12}
\multirow{2}{*}{M5V$^\dagger$} & 3055 	& 0.16 & 6 &	8.3		& 1.3 	& 220		& 3.6		&  9		& 23 & 60 & 158  \\
 	& 2920 	& 0.14 & 8.5 &	6.4		& 1.1 	& 286	 	& 2.2		& 5 		& 14 & 36 & 94   \\
\hline
\end{tabular}
\caption{  \footnotesize Integration times (in units of ``number of transits") needed to obtain the specified SNR and spectral resolution for a given stellar type/brightness (in Mag. K) in the MIR.
$\tau_{transit}$ is the transit duration given in hours, \textit{ph} stands for photometry, where a few wavelenghts can be probed for the most challenging targets. \emph{$\dagger,*$: See Table \ref{tab:189primary} caption.}}
   \label{tab:mainresults320}
\end{table}

\section{Discussion}
\subsection{Stellar Variability}

Our simulations do not include the effects of stellar variability on transit observations. Kepler is reaching 200 ppm/min on an V=11 mag star and 40 ppm/min on a V=7 mag star. The most up-to-date information about variability comes from the studies of \citet{basri2010,basri2011} based on the analysis of 100,000 stars (first release of 43 days of Kepler data). For timescales between 3-16 days, the authors showed that 57 \% of G stars are active and tend to be more active than the Sun (up to twice the activity level is typical). This fraction increases to 87\% of K and M dwarfs (figure 4 of \citet{basri2010}). The peak of the histogram of amplitude distribution is centered at 2 mmag. Scatter plots from Basri et al. show that for K and M stars indeed the dominant source of scatter is variability, not Poisson noise. The bulk of the periodicities is found at periods larger than 10 days, with amplitudes ranging from 1-10 mmag. \citet{ciardi2011} found that 80\% of M dwarfs have dispersion less than 500 ppm over a period of 12 hours, while G dwarfs are the most stable group down to 40 ppm.

It is important to note here that the photometric variability is significantly lower in the near infrared than in the Kepler band \citep{Agol2010,knutson2011}, because of the lower contrast between spots and the stellar photosphere at larger wavelengths. For instance, \citet{Agol2010} measured that the infrared flux variations in the case of the active K star HD 189733 are about 20\% of the optical variations. This is in agreement with the theoretical estimates by \citet{Ballerini2011}

Most importantly, all the timescales related to stellar activity patterns are very different from the timescales associated to single transit observations (a few hours), and thus can be easily removed. CoRoT-7 b provides a good example. The activity modulations are of the order of 2\% and yet CoRoT managed to find a transit with a depth of 0.03\%. This was made possible by the continuous monitoring provided by CoRoT and the different timescale compared with the transit signal that allowed for the removal of the activity effects and the discovery of variations smaller than the overall modulation by a factor of 70. The same situation has been encountered in the list of 1200 Kepler candidates announced recently, in which stellar activity modulations and transit events have been disentangled, often with the former being far greater than the latter.

In conclusion, the overall (random) photometric jitter of the star should not be a crucial factor with the right strategy to adequately correct for modulations caused by spot variations. Time series can be used as an ``activity monitor''  by the visible part of the spectrum.
As mentioned in 2.4, systematic differences in the stellar flux could hamper multiple transit combinations. However, where primary transit observations are subject to these effects, secondary eclipse observations are preferred as they are immune to them.

\subsection{Planetary Variability}

Upper limits about eclipse variability have been reported by \citet{Agol2010} and \citet{knutson2011}. We do not know the nature of this variability, but the chance of observing multiple 
spectra rather than photometric bands might be helpful to explore the potential sources of atmospheric variability (thermal changes? chemical changes? clouds/hazes?) 
for the most favorable targets.
In the case of faint targets, for which co-adding eclipse observations is necessary, only spatially/temporally-averaged information will be available. From the experience with the planets in our own Solar System, this information, although more limited, is expected to be still very significant.

\subsection{Stellar Population}

The integration times required to study habitable-zone super-Earths (given in table \ref{tab:mainresults320}) show that characterisation of these targets is possible provided they orbit late type dwarfs. While bright targets are preferred, as they provide a higher photon signal, our results cover a range of magnitudes from K=5 to K=9.
In parallel, the M type population found in the RECONS catalogue \citep{recons100}, which lists 100 stars up to 6.6pc in the Sun's local neighbourhood,
is mostly formed of bright targets with a significant fraction having magnitudes between K=4 and K=6 (see Fig. \ref{fig:recons}).
Extrapolation from the catalogue up to magnitude K=9 yields however a much larger stellar population that can be studied for super-Earths.
Thus, combining the feasibility of studying targets up to K=9, while keeping a preference for brighter sources, and the greater amount of fainter stars up to mag. K=9, creates a common area ideal for super-Earth observations centered around the K=7-8 magnitude region. A mission that aims to characterise habitable-zone super-Earths should have detectors optimised for this magnitude range.

\subsection{Instrument Transmission}
\label{sec:instru_discussion}
Throughout this paper we have considered an instrumental transmission value of 0.7. In practical applications, many factors can reduce this transmission value. While most of the cases presented allow for slightly longer observations, the most challenging category of habitable-zone super-Earths will require high instrumental transmission values to remain feasible. Instrument designs with high levels of transmission, such as fourier transform spectrographs, can be considered a possibility for the characterisation of these most challenging targets. 

\subsection{Systematic Effects}

We presented here idealised cases where systematic errors (such as detector time constants, pointing jitter, re-acquisition errors, temperature fluctuations, etc.) were not accounted for. 
Instrumental settings for our results from the visible to the infrared were based on available technology and can be considered realistic. 
With these considerations, the results presented in this paper highlight that in the coming years habitable-zone super-Earths are realistically within reach. In future work, we will update our models as information on the systematic effects of specific instruments becomes available.
\section{Conclusions}
We have presented in this paper a detailed study of the performances and trade-offs of a M-class transit spectroscopy mission dedicated to the observation of exoplanetary atmospheres.
We have demonstrated that, in principle, with a 1.2/1.4\,m space telescope performing simultaneous spectroscopy from the visible to the mid-IR, we are able to secure the characterisation of a plethora of exoplanets, ranging from the hot, gaseous down to the temperate ones approaching the size of the Earth. 
According to our simulations, the spectra of hot-Jupiters orbiting F, G and K-type stars with V mag. brighter than 10 can be obtained by integrating from a fraction of transit up to few tens of transits to reach a spectral resolution of 300 and SNR = 50. 
Habitable-zone super-Earths are undoubtedly the most challenging category of targets due to their small size, low temperature and their relatively large separation from the star. We show however, that these targets can be observed at low resolution in the Mid-IR, provided their hosting star is a bright M dwarf. While most of the Sun's neighbourhood is composed of these late-type stars, efforts still need to be directed at increasing the number of low mass stars known and constraining their properties. The 2MASS catalogue sample, completed with current and planned dedicated ground-based surveys, as well as space missions such as  WISE and GAIA should offer a viable solution to this critical issue in the next five years. \\
In future work, we will update our current instrument models by including  a more realistic treatment of the systematics.

\section{Acknowledgements}
We wish to thank Laura Affer for her work on analysing 2MASS data through color-color diagrams, Ingo Waldmann for discussions on signal-to-noise issues, and in particular the anonymous referee for the helpful comments.
\appendix
\section{Appendix}
In addition to the numbers presented throughout the paper for a 1.4m telescope, we provide here two supplementary sets of results for a 1.2m telescope.
We detail in Table \ref{tab:appendixinfo} the parameters adopted for the two cases. The results are displayed in the following way: Number of transits: Case 1 \textit{(Case 2)}.
\begin{table*}[h]
\footnotesize
\centering
\begin{tabular}{ l  c  c  c }
 &\multicolumn{2}{c}{Case 1} & Case 2\\
\hline
Detector used & \multicolumn{2}{c}{SOFRADIR}  & RAYTHEON\\
&  LWIR & VLWIR & JWST Si:As\\
\hline
Spectral range considered ($\mu$m) &  5 - 11  & 11 - 16 & 5 - 16\\
\hline

Full well capacity (electrons) &  $2 \cdot 10^7$ & $5 \cdot 10^6$ & $2 \cdot 10^5$ \\

Dark current (electrons/s/pixel) &  500 & 300 & 0.2 \\

Quantum efficiency (electrons/photon)&   0.7 & 0.7 & 0.7\\

Readout noise (electrons/pixel/readout) & 1000 & 1000 & 15 \\

Readout time (seconds) &  0.03 & 0.01 & 3 \\
\hline
Telescope temperature (K) &  $<60$ & $<60$ & $<60$ \\

Instrument temperature (K) &  45 & 45 &  45 \\

Telescope transmission & 0.9 & 0.9  & 0.85\\

Instrument transmission & 0.7 & 0.7 & 0.4\\
\hline
\end{tabular}
\caption{\footnotesize List of parameters used in the two sets of appendix results. In the first case, two detectors are needed to cover the 5 to 16 micron range, while for the second set of results, which represents an alternate design of the instruments, one detector is used for the full range. The results are split into four columns representing wavelength bands used. The first column lists values in the photometric N band, which is also the band used for results presented throughout the paper, followed by three channels: 5 to 8.3 $\mu m$, 8.3 to 11 $\mu m$ and 11 to 16 $\mu m$. A 30 $\mu$m pixel size and 2 illuminated pixels per spectral element are assumed (For the N band (7.7 to 12.7 $\mu$m) we have used the LWIR setting values). In the case of the VLWIR detector, we have used a dark current value of 300 electrons/s/pixel considering existing technologies and expected future capabilities. Further discussion on these values can be found in section \ref{sec:instru_discussion}.
 }
\label{tab:appendixinfo}
\end{table*}
\newpage

\subsection{1.2m telescope, Hot Planets}
\footnotesize

 \begin{center}
\small
\begin{tabular}{  c  c  c  c  c }
\hline
Bands:  &  N (7.7 to 12.7) &  5 to 8.3  &  8.3 to 11  &  11 to 16 \\
1) Contrasts: &1.01E-03 & 5.13E-04 & 8.34E-04 & 7.21E-04\\
\hline
V=5 &     9.56  \textit{(15.71)} &     12.62  \textit{(21.22)} &     13.60  \textit{(22.38)} &     41.05  \textit{(58.11)}\\
V=6 &     25.29 \textit{(39.49)} &     32.30  \textit{(53.31)} &     35.94  \textit{(56.21)} &     111.60  \textit{(157.43)}\\
V=7 &     71.63 \textit{(99.33)} &     84.94  \textit{(133.92)} &     101.47  \textit{(141.22)} &     LR  \textit{(LR)}\\
V=8 &     LR \textit{(LR)} &    LR  \textit{(LR)} &    LR  \textit{(LR)} &   LR  \textit{(LR)}\\
V=9 &     LR \textit{(LR)} &    LR  \textit{(LR)} &     LR  \textit{(LR)} &   LR  \textit{(LR)}\\
\hline
2) Contrasts: &5.56E-03 & 2.89E-03 & 4.61E-03 & 3.93E-03\\
\hline
V=5 &     0.21 \textit{(0.36)} &     0.27 \textit{(0.50)} &     0.30 \textit{(0.50)} &     0.90 \textit{(1.28)} \\
V=6 &     0.54 \textit{(0.89)} &     0.67 \textit{(1.13)} &     0.77 \textit{(1.26)} &     2.35 \textit{(3.33)} \\
V=7 &     1.44 \textit{(2.24)} &     1.73 \textit{(2.84)} &     2.04 \textit{(3.18)} &     6.42 \textit{(9.05)} \\
V=8 &     4.12 \textit{(5.64)}&     4.55 \textit{(7.14)} &     5.80 \textit{(7.98)} &    19.33 \textit{(27.07)} \\
V=9 &     13.42 \textit{(14.23)}&     12.76 \textit{(17.95)} &    18.81 \textit{(20.06)} &    68.83 \textit{(95.35)} \\
\hline
3) Contrasts: &1.38E-04 & 8.61E-05 & 1.32E-04 & 1.69E-04\\
\hline
K=5 &    17.86 \textit{(30.06)} &    15.30 \textit{(36.55)} &    19.15 \textit{(32.23)} &    25.47 \textit{(36.29)} \\
K=6 &    45.71 \textit{(75.54)} &    38.68 \textit{(65.21)} &    48.97 \textit{(80.96)} &    66.87 \textit{(95.81)} \\
K=7 &    120.18 \textit{(189.99)} &    98.68 \textit{(163.79)} &    128.52 \textit{(203.40)} &   186.32 \textit{(270.04)} \\
K=8 &   335.66 \textit{(478.83)} &   257.50 \textit{(411.44)} &   357.59 \textit{(511.12)} &   583.75 \textit{(863.62)} \\
K=9 &   1056.31 \textit{(1212.80)} &   707.50 \textit{(1033.57)} &   1117.56 \textit{(1285.18)} &  LR \textit{(LR)} \\
\hline
4) Contrasts: &1.22E-03 & 7.78E-04 & 1.17E-03 & 1.48E-03\\
\hline
K=5 &     0.63 \textit{(1.06)} &     0.51 \textit{(1.23)} &     0.67 \textit{(1.13)} &     0.90 \textit{(1.29)} \\
K=6 &     1.61 \textit{(2.66)} &     1.28 \textit{(2.16)} &     1.72 \textit{(2.84)} &     2.73 \textit{(3.40)} \\
K=7 &     4.23 \textit{(6.69)} &     3.27 \textit{(5.44)} &     4.51 \textit{(7.13)} &     6.60 \textit{(9.56)} \\
K=8 &     11.82 \textit{(16.87)} &     8.54 \textit{(13.65)} &     12.54 \textit{(17.92)} &    20.63 \textit{(30.52)} \\
K=9 &    37.19 \textit{(42.72)} &    23.42 \textit{(34.30)} &    39.18 \textit{(45.07)} &    77.42 \textit{(117.64)} \\
\hline
\end{tabular}
\captionof{table}{ \footnotesize 1: Integration times in number of transits for a hot Jupiter orbiting a F3.0V star. The four columns compare integration times in different bands for the same target. The contrast value and number of resolution elements are given for each band. The five rows list results for the specified star with varying magnitude (here in mag. V). The star temperature used is 6740 K, and the transit duration assumed is 2.90 hours. A spectral Resolution of 300 and a SNR value of 50 are used. A dash `-' signifies that the number of transits required is over the maximum number of transits that can be covered over a mission lifetime. `LR' stands for Lower Resolution, and is indicated when observations need to be done at a lower spectral resolution to fit within the time constrains of a mission, and `phot' stands for photometry at selected wavelengths, where lower resolution is not feasible.\\
2: Planet: Hot Jupiter, Star: K1V, temp: 4900K, R=300,  SNR=50.\\
3: Planet: Hot SE, Star: M1.5V, temp: 3582K, R=40,  SNR=10. \\
4: Planet: Hot SE, Star: M5V, temp: 3055K R=40,  SNR=10.  }
\label{tab:appendix1}
\end{center}

\subsection{1.2m telescope, Warm Planets}

\begin{center}
\centering
\small
\begin{tabular}{  c c c c c }
\hline
Bands:  &  N (7.7 to 12.7) &  5 to 8.3  &  8.3 to 11  &  11 to 16 \\
1) Contrasts: &4.61E-04 & 3.10E-04 & 4.10E-04 & 1.28E-03\\
\hline
K=5 &    19.39 \textit{(32.52)} &    14.12 \textit{(27.06)} &    23.82 \textit{(39.97)} &     5.31 \textit{(7.57)} \\
K=6 &    49.84 \textit{(81.74)} &    35.75 \textit{(60.11)} &    61.18 \textit{(100.40)} &    14.03 \textit{(20.07)} \\
K=7 &    132.40 \textit{(205.61)} &    91.55 \textit{(151.00)} &   162.21 \textit{(252.24)} &    39.56 \textit{(57.18)} \\
K=8 &   378.10 \textit{(518.31)} &   241.00 \textit{(379.32)} &   461.30 \textit{(633.88)} &    126.66 \textit{(186.27)} \\
K=9 &   LR \textit{(LR)} &   675.07 \textit{(LR)} &   LR \textit{(LR)} &   490.38 \textit{(LR)} \\
\hline
2) Contrasts: &1.93E-04 & 7.12E-05 & 1.75E-04 & 2.94E-04\\
\hline
K=5 &     5.55 \textit{(9.40)} &     13.49 \textit{(65.35)} &     6.62 \textit{(11.22)} &     5.06 \textit{(7.22)} \\
K=6 &    14.08 \textit{(23.63)} &    34.00 \textit{(65.35)} &    16.79 \textit{(28.18)} &     13.13 \textit{(18.86)} \\
K=7 &    36.22 \textit{(59.41)} &    86.06 \textit{(144.73)} &    43.14 \textit{(70.78)} &    35.62 \textit{(51.91)} \\
K=8 &    96.34 \textit{(149.68)} &   220.38 \textit{(363.54)} &    114.42 \textit{(177.86)} &    106.09 \textit{(159.04)} \\
K=9 &   275.81 \textit{(378.73)} &  580.07 \textit{(LR)} &   325.63 \textit{(447.17)} &  371.25 \textit{(580.26)} \\
\hline
3) Contrasts: &3.29E-04 & 1.22E-04 & 2.98E-04 & 4.98E-04\\
\hline
K=5 &     2.46 \textit{(4.17)} &     5.86 \textit{(28.50)} &     2.93 \textit{(4.96)} &     2.25 \textit{(3.21)} \\
K=6 &     6.24 \textit{(10.47)} &    14.76 \textit{(28.51)} &     7.43 \textit{(12.47)} &     5.84 \textit{(8.39)} \\
K=7 &    16.05 \textit{(26.33)} &    37.36 \textit{(62.82)} &    19.09 \textit{(31.32)} &    15.84 \textit{(23.08)} \\
K=8 &    42.69 \textit{(66.33)} &    95.65 \textit{(157.81)} &    50.62 \textit{(78.69)} &    47.15 \textit{(70.67)} \\
K=9 &    122.22 \textit{(167.84)} &   251.71 \textit{(396.40)} &   144.07 \textit{(197.85)} &   164.88 \textit{(257.66)} \\
\hline
\end{tabular}
\captionof{table}{ \footnotesize See Table \ref{tab:appendix1} for additional explanation.\\1: Planet: Warm Neptune, Star: M2.5V, temp: 3480K, R=50,  SNR=30.\\
2: Planet: Warm SE, Star: M4V, temp: 3230K, R=20,  SNR=10.\\
3: Planet: Warm SE, Star: M5V, temp: 3055K, R=20,  SNR=10.\\
}
\label{tab:appendix10}
\end{center}

\subsection{1.2m telescope, HZ Planets}

\begin{center}
\centering
\small
\begin{tabular}{  c c c c c }
\hline
Bands:  &  N (7.7 to 12.7) &  5 to 8.3  &  8.3 to 11  &  11 to 16 \\
1) Contrasts: &1.53E-04 & 2.12E-06 & 1.27E-04 & 1.58E-04\\
\hline
V=5 &     0.35 \textit{(1.86)}  &   phot \textit{(-)} &     0.49 \textit{(2.70)}  &     0.69 \textit{(1.75)} \\
V=6 &     0.87 \textit{(1.86)}  &  - \textit{(-)} &     1.24 \textit{(2.70)}  &    1.74 \textit{(2.47)} \\
V=7 &     2.21 \textit{(3.72)} &  - \textit{(-)} &     3.12 \textit{(5.26)}  &     4.44 \textit{(6.32)} \\
V=8 &     5.65 \textit{(9.36)} &  - \textit{(-)} &     7.98 \textit{(13.22)} &     11.63 \textit{(16.66)} \\
V=9 &    14.83 \textit{(LR)} & - \textit{(-)} &    LR \textit{(LR)} &    LR \textit{(LR)} \\
\hline
2) Contrasts: &3.54E-05 & 4.97E-06 & 2.89E-05 & 8.15E-05\\
\hline
K=5 &     11.60 \textit{(36.69)} &   phot \textit{(-)} &    16.91 \textit{(59.39)} &   4.60 \textit{(7.84)} \\
K=6 &    29.28 \textit{(52.81)} &   phot \textit{(-)} &    42.68 \textit{(76.96)} &     11.87 \textit{(18.20)} \\
K=7 &    74.47 \textit{(132.80)} &   - \textit{(-)} &    108.43 \textit{(phot)} &    31.75 \textit{(49.50)} \\
K=8 &   phot \textit{(-)} &  - \textit{(-)} &   phot \textit{(-)} &    92.00 \textit{(phot)} \\
\hline
3) Contrasts: &8.46E-05 & 1.21E-05 & 6.92E-05 & 1.93E-04\\
\hline
K=5 &     2.95 \textit{(10.42)} &    47.51 \textit{(-)} &     4.29 \textit{(15.55)} &     1.18 \textit{(2.08)} \\
K=6 &     7.46 \textit{(13.87)} &    119.51 \textit{(-)} &     10.83 \textit{(20.15)} &     3.04 \textit{(4.81)} \\
K=7 &    18.96 \textit{(34.87)} &   phot \textit{(-)} &    27.53 \textit{(50.61)} &     8.13 \textit{(13.07)} \\
K=8 &    49.10 \textit{(87.83)} &   phot \textit{(-)} &    71.12 \textit{(127.18)} &    23.53 \textit{(39.10)} \\
K=9 &   132.62 \textit{(222.11)} &  - \textit{(-)} &   191.12 \textit{(phot)} &    78.75 \textit{(137.79)} \\
\hline
\end{tabular}
\captionof{table}{ \footnotesize See Table \ref{tab:appendix1} for additional explanation.\\1: Planet: HZ Jup, Star: K4V, temp: 4780K, R=40,  SNR=10.\\
2: Planet: HZ SE, Star: M4V, temp: 3230K, R=10,  SNR=5.\\
3: Planet: HZ SE, Star: M5.5V, temp: 2920K, R=10,  SNR=5.
}
\label{tab:appendix15}
\end{center}

   \bibliographystyle{apalike}       
\bibliography{tess3110}
\end{document}